\documentclass[conference]{IEEEtran}
\IEEEoverridecommandlockouts
\usepackage{cite}
\usepackage{amsmath,amssymb,amsfonts}
\usepackage{textcomp}
\usepackage[utf8]{inputenc} % allow utf-8 input
\usepackage[T1]{fontenc}    % use 8-bit T1 fonts
\usepackage{hyperref}       % hyperlinks
\usepackage{url}            % simple URL typesetting
\usepackage{booktabs}       % professional-quality tables
\usepackage{amsfonts}       % blackboard math symbols
\usepackage{nicefrac}       % compact symbols for 1/2, etc.
\usepackage{microtype}      % microtypography
\usepackage{xcolor}         % colors
\usepackage{multirow}
\usepackage{graphicx}
\usepackage{amsmath}
\usepackage{diagbox}
\usepackage{multirow}
\usepackage{amsfonts}
\usepackage{subfigure}
\usepackage{amsmath}
\usepackage{diagbox}

\usepackage{enumerate}
\usepackage{xcolor}
\usepackage{array}
\usepackage{balance}
\usepackage{amssymb}
\usepackage{mathrsfs}
\usepackage{bm}
\usepackage{algorithm}
\usepackage{algorithmic}
\usepackage{multirow}
\usepackage{soul}
\soulregister{\cite}{7}
\soulregister{\ref}{7}
\soulregister{\item}{7}
\soulregister{\textbf}{7}
\usepackage{tikz}
\usetikzlibrary{bayesnet}
\tikzset{fontscale/.style = {font=\relsize{#1}}}

\def\BibTeX{{\rm B\kern-.05em{\sc i\kern-.025em b}\kern-.08em
		T\kern-.1667em\lower.7ex\hbox{E}\kern-.125emX}}
\begin{document}
	
	\title{Diverse Preference Augmentation with Multiple Domains for Cold-start Recommendations
%		\thanks{$^*$Corresponding authors: Lixin Duan, Hui Xu.}
%		{This work was supported by the Major Project for New Generation of AI (grant No. 2018AAA0100400), the National Natural Science	Foundation of China (grants No. 61832001 and No. 62002052), Sichuan	Science and Technology Program (grant No. 2021JDRC0079), the China Postdoctoral Science Foundation (No. 2019TQ0051) and the Australian Research Council (grants No. DP180100106 and No. DP200101328).}
	}
	
 	\author{
 		\IEEEauthorblockN{Yan Zhang$^{\dagger \ddagger \natural}$, Changyu Li$^{\dagger}$, Ivor W. Tsang$^{\sharp \ddagger}$, Hui Xu$^{\dagger }$, Lixin Duan$^{\dagger }$, Hongzhi Yin$^{\mathsection}$, Wen Li$^{\dagger}$, Jie Shao$^{\dagger }$}
 		\IEEEauthorblockA{$^{\dagger}$University of Electronic Science and Technology of China, Chengdu 611731, China \\
 			$^\ddagger$Australian Artificial Intelligence Institute, University of Technology Sydney, Sydeny 2007, Australia \\	
 	    	$^\natural$Intelligent Terminal Key Laboratory of Sichuan Province, Yibin 644000, China \\		
 			$^\sharp$Center for Frontier AI Research, Research Agency for Science, Technology and Research (A$^\star$STAR), Singapore \\	
% 			\textit{$^\natural$Sichuan Artificial Intelligence Research Institute}, Yibin 644000, China \\
 			$^{\mathsection}$The University of Queensland, Brisbane 4067, Australia \\
 			\textit{yixianqianzy@gmail.com, changyulve@std.uestc.edu.cn, Ivor.Tsang@gmail.com, hui\_xu@std.uestc.edu.cn,}\\
 		\textit{lxduan@uestc.edu.cn, db.hongzhi@gmail.com, \{liwen, shaojie\}@uestc.edu.cn}}}
	%\and
	%		\IEEEauthorblockN{}
	%	\IEEEauthorblockA{\textit{School of Computer Science and Engineering} \\
	%		\textit{University of Electronic Science and Technology of China}\\
	%		Chengdu, China \\
	%		}
	%\and
	%	\IEEEauthorblockN{Hui Xu}
	%		\IEEEauthorblockA{\textit{School of Computer Science and Engineering} \\
	%		\textit{University of Electronic Science and Technology of China}\\
	%		Chengdu, China \\
	%		}
	%\and
	%	\IEEEauthorblockN{Lixin Duan}
	%	\IEEEauthorblockA{\textit{School of Computer Science and Engineering} \\
	%		\textit{University of Electronic Science and Technology of China}\\
	%		Chengdu, China \\
	%		}
	%\and
	%	\IEEEauthorblockN{Jie Shao}
	%	\IEEEauthorblockA{\textit{School of Computer Science and Engineering} \\
	%		\textit{University of Electronic Science and Technology of China}\\
	%		Chengdu, China \\
	%		shaojie@uestc.edu.cn}
	%	
	%}

	\maketitle
	
	\begin{abstract}
		Cold-start issues have been more and more challenging for providing accurate recommendations with the fast increase of users and items. Most existing approaches attempt to solve the intractable problems via content-aware recommendations based on auxiliary information and/or cross-domain recommendations with transfer learning. Their performances are often constrained by the extremely sparse user-item interactions, unavailable side information,  or very limited domain-shared users. Recently, meta-learners with meta-augmentation by adding noises to labels have been proven to be effective to avoid overfitting and shown good performance on new tasks. Motivated by the idea of meta-augmentation, in this paper, by treating a user's preference over items as a task, we propose a so-called Diverse Preference Augmentation framework with multiple source domains based on meta-learning (referred to as MetaDPA) to i) generate diverse ratings in a new domain of interest (known as target domain) to handle overfitting on the case of sparse interactions, and to ii) learn a preference model in the target domain via a meta-learning scheme to alleviate cold-start issues. Specifically, we first conduct multi-source domain adaptation by dual conditional variational autoencoders and impose a Multi-domain InfoMax (MDI) constraint on the latent representations to learn domain-shared and domain-specific preference properties. To avoid overfitting, we add a Mutually-Exclusive (ME) constraint on the output of decoders to generate diverse ratings given content data. Finally, these generated diverse ratings and the original ratings are introduced into the meta-training procedure to learn a preference meta-learner, which produces good generalization ability on cold-start recommendation tasks. Experiments on real-world datasets show our proposed MetaDPA clearly outperforms the current state-of-the-art baselines.
	\end{abstract}
	
	\begin{IEEEkeywords}
		Recommender system, preference augmentation, meta leaning, cold-start
	\end{IEEEkeywords}
	
	\section{Introduction}\label{sec_intro}
	Recommender systems have shown great success in both academia and industries, and so become indispensable in our life by helping us filter millions of possible choices. Recommender systems provide a small set of items from the underlying pool of items based on users’ historical interactions and their side information. One of the well known recommendation frameworks is Collaborative Filtering (CF)\cite{schafer2007collaborative, he2017neural}, where the only available data is user-item historical interactive information. A key challenge in CF-based methods is to provide accurate recommendations from a large number of items with extremely sparse interactions\cite{zou2020neural}. Such recommender systems suffer from poor performance due to sparse interactive data or ratings and cannot even handle user cold-start and item cold-start issues brought by new users and new items \cite{lee2019melu}. So as a very critical problem in recommender systems, how to make accurate recommendation under sparse and cold-start scenarios attracts rising attention from a wide range of stakeholders in recent years.

	Existing approaches for solving cold-start and sparse issues are proposed from the following three directions: (1) content-aware recommender systems\cite{wang2011collaborative,wang2015collaborative,agarwal2009regression,rendle2012factorization,lian2015content} that integrate auxiliary side information and interactive data to enhance representations of users and items, and then feed them into the preference model to improve the performance of recommender systems, where user's and item's content information (user's profile, item's description, user's review, etc.) is taken as auxiliary side information for strengthening representations of users and items. (2) Cross-domain recommender systems transfer preference knowledge from source domains to its similar target domain, and then improve the performance of recommendations in the target domain. These methods can be categorized into single source\cite{zhao2017collaborative,chen2020multi,ding2019learning,nguyen2017personalized,he2018robust,guo2021gcn,cantador2015cross,lian2017cccfnet} and multiple source\cite{singh2008relational,cortes2018cold,kim2019domain,zhang2012multi,yi2018multi,yan2019multi,zhang2015personalization} cross-domain recommendations based on the number of source domains applied for transferring preference knowledge. 
	%Single source cross-domain recommendations\cite{zhao2017collaborative,ding2019learning,nguyen2017personalized,he2018robust,cantador2015cross,lian2017cccfnet} transfer preference knowledge from one source domain to the target domain. Multi-source cross-domain recommendations\cite{singh2008relational,cortes2018cold,kim2019domain,zhang2012multi,yi2018multi,yan2019multi,zhang2015personalization} transfer preference properties from multiple source domains to the target domain. 
	(3) Meta-learning based recommender systems\cite{DBLP:conf/nips/VartakTMBL17,DBLP:conf/kdd/DongYYXZ20,DBLP:conf/kdd/Lu0S20,DBLP:conf/nips/VartakTMBL17} learn the prior preference distribution of users over items by taking users' preferences as meta-learning tasks. They can improve the performance of recommendations under sparse and cold-start settings by fine-tuning the preference model with only a few ratings.
	
	Content-aware recommender systems have been widely studied for solving the cold-start and sparse issues with the help of auxiliary side information. One of the most commonly used side information is content\cite{wang2011collaborative,wang2015collaborative,agarwal2009regression,rendle2012factorization,lian2015content,he2017neural,zhang2019deep,he2017neural2,xue2017deep}. By learning with content data, the features of users or items with just a few ratings or even no ratings can be then effectively represented, making those content-aware recommender systems be able to improve the performance under sparse and cold-start scenarios. However, there exist inconsistencies between item content and user preferences. That is to say, users under the same profile (e.g., age, occupation, gender, place of residence, etc.) often have different preferences over the same item; and items having similar content (e.g., description, category, users’ reviews, etc.) are often rated with different scores by the same user. As a result, the performance improvement of
content-aware recommender systems are limited by the gap between content and preferences.
	
	Cross-domain recommendations based on transfer learning are another type of solutions for solving cold-start and sparse challenges. Existing methods transfer knowledge from source domains with rich preference information to a target domain with very sparse historical data\cite{zhao2017unified,ding2019learning,nguyen2017personalized,he2018robust}. Moreover, those methods transfer preference information by domain adaptation with domain-shared users in order to strengthen the representations of users and items. Thus, the cross-domain model can learn effective representations of users and items than only using preference information of a single domain. However, the limit number of shared users affects the capability of preference transferring, which restricts the performance improvement in the target domain. For example, on Amazon datasets, Books and Electronics subsets only share 5\% users, which limits the transferable preference patterns from the source domain to the target domain, and thus the performance of cross-domain recommendations is also constrained.	
	
	To acquire more preference patterns, some scholars study multi-domain recommendations of transferring preference patterns from multiple source domains to a target domain\cite{singh2008relational, cortes2018cold, kim2019domain, zhang2012multi, yi2018multi, yan2019multi, zhang2015personalization}.  These methods extract correlations between source domains and the target domain and tie factors from different source domains together. Such correlation enrich rating patterns of the target domain with multiple related source domains. Thus, these methods can achieve better performance than single source domain. The transferable preference patterns are also limited because the correlation is dependent on shared users. The performance is still limited by the ratio of shared users among all users. In addition, some of these multi-source cross-domain methods can only provide recommendation for shared users and do not work for providing recommendations for unshared users. Besides, the augmentation method is another way to acquire more preference patterns, such as AugCF\cite{wang2019enhancing}. It is designed based on Conditional Generative Adversarial Nets by considering the class (like or
dislike) as a feature to generate new interaction data, which is evaluated to be
a sufficiently real augmentation to the original dataset in their work.
	
	Meta-learning has been validated as a promising approach for mitigating cold-start issues of recommendations\cite{lee2019melu,dong2020mamo,lin2021task, DBLP:conf/kdd/Lu0S20, DBLP:conf/nips/VartakTMBL17}, which treats user's preferences as meta-learning tasks and learns a preference prior distribution of all users over items. Meta-learning based methods can fast adapt to new users' or new items' recommendations with the learned meta-learner of the preference prediction model. However, the meta-learner easily overfits to the sparse preference (rating) data, which makes it difficult to provide accurate recommendations under new users or new items settings. 
	
   To avoid overfitting on meta-training tasks, meta-augmentation\cite{rajendran2020meta} adds noise to labels $y$ without changing inputs $x$. It is capable of handling two forms of overfitting: (1) memorization overfitting, in which the model is able to overfit to the training set without relying on the meta-learner, and (2) meta-learner overfitting, in which the learner overfits to the training set and does not generalize to the test set. Yin et al.\cite{yin2019meta} identify the memorization overfitting can happen when the set of tasks are non-mutually-exclusive. The meta-augmentation is proven to avoid memorization overfitting effectively by transforming the task setting from non-mutually-exclusive to mutually-exclusive and proven to avoid the learner overfitting effectively\cite{yin2019meta}. Tasks are said to be mutually-exclusive\cite{rajendran2020meta} if a single model cannot solve them all at once. For example, if the task $\mathcal{T}_1$ is `output 0 if the input image is a dog', and task $\mathcal{T}_2$ is `output 1 if the image is a dog', then we call tasks $\{\mathcal{T}_1, \mathcal{T}_2\}$ are mutually-exclusive. 
   
   Generally, we define \textit{`Mutual Exclusivity'} as: Training samples $(x, y_{1}), (x, y_{2}), \cdots, (x, y_{k})$ are called mutually-exclusive samples, if all continuous labels $y_{1}, y_{2}, \cdots, y_{k}$ are different from each other with the same input $x$. For example, training samples $(x, 0.1), (x, 0.2), (x, 0.3)$ with the same input $x$ are mutually-exclusive samples because all labels are different from each other. However, in practice, it is difficult to obtain training samples that meet such a strict assumption (i.e., all labels are required to be different with the same input). By relaxing the assumption, we define \textit{`Diversity'} as: Training samples $(x, y_{1}), (x, y_{2}), \cdots, (x, y_{k})$ are called diverse samples, if not all continuous labels $y_{1}, y_{2}, \cdots, y_{k}$ are different from each other with the same input $x$. For example, training samples $(x, 0.1), (x, 0.1), (x, 0.3)$ are diverse samples because not all labels are different (the first and the second samples are with the same label $0.1$). It's worth noting that labels in the above two definitions are continuous values. Particularly, continuous labels are within the interval $[0, 1]$, because we train our model on real interactive data(`0' or `1') and augmented interactive data within the interval $[0, 1]$. It is also worth mention that all labels are required to be different from each other in mutually-exclusive samples; while in diverse samples, some labels from $\{y_{1}, y_{2}, \cdots, y_{k}\}$ may be the same, so mutually-exclusive samples are diverse samples, but not vice versa.
   %For example, training samples $(x, 0.1), (x, 0.2), (x, 0.3)$ with the same input $x$ are called as mutual exclusive samples; and }
   	%For example, in training samples $(x, y_{1}), (x, y_{2}), (x, y_{3})$ with the same input $x$, if $y_{1}=0$, $y_{2}=0.5$ and $y_{3}=1$, i.e., labels $y_{1} \neq y_{2}$, $y_{2} \neq y_{3}$ and $y_{1} \neq y_{3}$, then we call $(x, y_{1}), (x, y_{2}), (x, y_{3})$ as mutually exclusive samples; if labels $y_{1}= y_{2} = 0$ and $y_{3}=0.5$, i.e., $y_{1} = y_{2} \neq y_{3}$, then we call them as diverse samples. }

%   But it is difficult to construct mutually exclusive tasks in real world due to the strong constraint. In the work\cite{rajendran2020meta}, it augments diverse tasks by the meta-augmentation via adding noises to labels $y$ without changing inputs $x$, i.e., it constructs tasks $\{(x, y+\epsilon_1), (x, y+\epsilon_2), \cdots\}$ via the task $(x, y)$, which aims to increase task diversity\cite{rajendran2020meta}. 

\begin{figure}[!htp]
		\centering
		\includegraphics[width=0.9\linewidth]{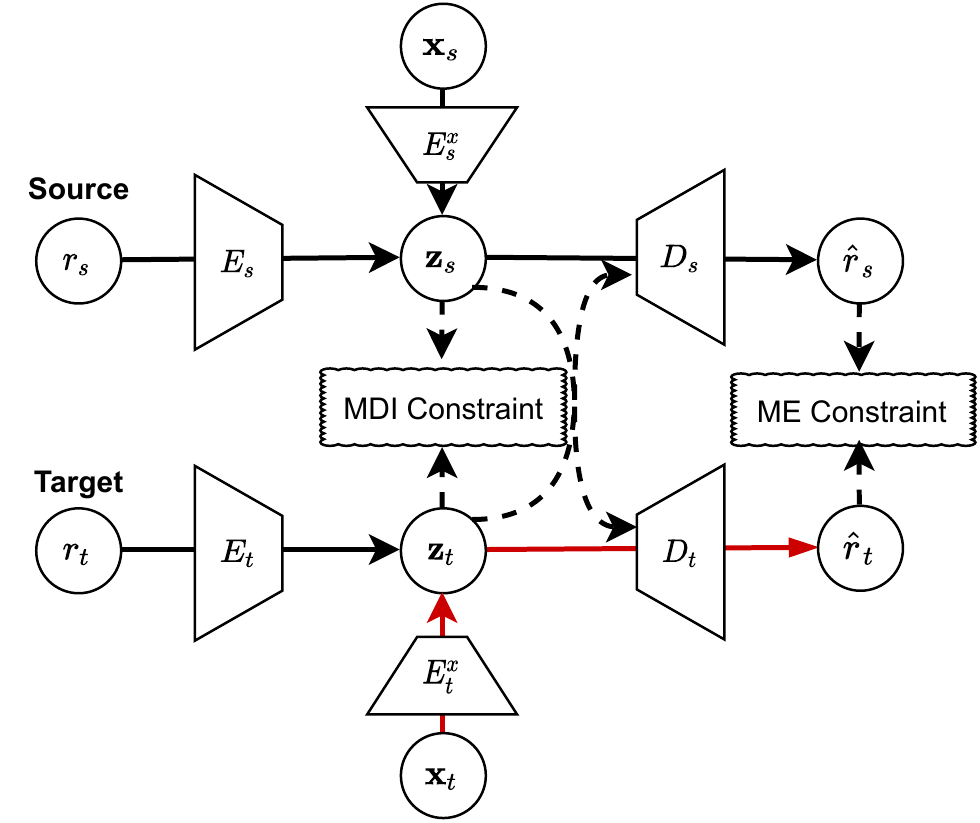}
			\vspace{-10pt}
		\caption{The framework of cross-domain adaptation with Dual-CVAE. The MDI constraint is imposed on latent representations $\mathbf{z}_s$ and $\mathbf{z}_t$. The ME constraint is enforced on the outputs of two decoders $D_s$ and $D_t$. Given ratings $\mathbf{r}_{s}$, $\mathbf{r}_t$ and content data $\mathbf{x}_{s}$, $\mathbf{x}_{t}$, the goal of Dual-CVAE is to learn domain-shared and domain-specific preference information, and then generate diverse ratings by decoders of $k$ Dual-CVAEs.}
		\vspace{-10pt}
		\label{fig:generation_network}
	\end{figure}

To avoid overfitting, our initial goal is to construct mutually-exclusive samples (tasks) in the target domain for training our model. Specifically, we develop a multi-source domain adaptation module that transfers the preference patterns from multiple source domains ($k$ source domains) to a target domain by $k$ dual conditional variational autoencoders (Dual-CVAEs) shown in Fig.~\ref{fig:generation_network}. By enforcing Mutually-Exclusive (ME) constraints on decoders of those $k$ Dual-CVAEs, we hope to generate mutually-exclusive ratings by $k$ encoder-decoder frameworks (highlighted with red line in Fig.~\ref{fig:generation_network}) given the input $\mathbf{x}_{t}$. 
%Therefore, we intend to learn $k$ Dual-CVAEs for capturing domain-specific information from $k$ different source domains by enforcing Mutually Exclusive (ME) constraints on decoders of these $k$ Dual-CVAEs, which is expected to generate mutually exclusive ratings from decoders of $k$ Dual-CVAEs in Fig.~\ref{fig:generation_network} (highlighted by red line). 
However, it is computationally intractable to obtain an optimal solution due to the strict assumption (i.e., all generated ratings from different source domains are required to be different from each other), because it needs $\mathcal{O}(k^2)$ operations to assert the assumption in each iteration of training those $k$ Dual-CVAEs. Accordingly, we relax the ME constraint by adding it to the objective with a weight (hyper-parameter $\beta_{2}$ in Eq. \eqref{dual}). Hence, these generated ratings may not be all different from each other, but they increase the rating diversity\cite{rajendran2020meta}. Consequently, we generate diverse ratings $\mathbf{\hat{r}}_{t1}, \mathbf{\hat{r}}_{t2}, \cdots, \mathbf{\hat{r}}_{tk}$ by the learned $k$ encoder-decoder frameworks with the same content $\mathbf{x}_{t}$ in the target domain, which is different from the work\cite{rajendran2020meta} where it augments diverse samples by adding noises to the ground-truth label $y$. 

We call our proposed method as Diverse Preference Augmentation based on meta-learning (MetaDPA). To learn domain-shared and domain-specific information by these $k$ Dual-CVAEs, we add Multi-domain InfoMax (MDI) constraints imposed on the latent representations of source and target domains\cite{federici2020learning}, which maximizes the mutual information between representations of source and target domains. To augment diverse ratings, we impose ME constraints on the decoders of $k$ Dual-CVAEs. Then we adopt the learned $k$ encoder-decoders (red line in Fig.~\ref{fig:generation_network}) to generate diverse ratings by the content data $\mathbf{x}_{u}$ of target domain. Finally we combine the augmented diverse ratings and true ratings to learn a prior preference distribution via a meta-learning framework in the target domain, which is expected to quickly adapt to new users’ or new items’ recommendations via only a few fine-tuning steps. The main contributions are summarized as follows:   	

	\begin{itemize}
		\item We propose a multi-source cross-domain recommender system, coined as Diverse Preference Augmentation based meta-learning (MetaDPA), for solving cold-start and sparse issues in recommendations. MetaDPA consists of three blocks: multi-source domain adaptation, diverse preference augmentation, and preference meta-learning. 
		\item For multi-source domain adaptation, MetaDPA augments diverse ratings with content data via the ME constraint imposed on the multi-source domain adaptation.
		\item We develop the MDI constraint to learn domain-shared and domain-specific preference information, where the domain-shared information makes it possible to transfer preference patterns from the source domain to the target domain, and the domain-specific preference information contributes to generate diverse ratings in the preference augmentation step.
		\item In experiments, we demonstrate the effectiveness of MetaDPA to relieve the sparse issue (`Warm-start') and three types of cold-start issues (`C-U', `C-I', `C-UI') by comparing with existing competitive baselines. Besides, we conduct ablation studies to evaluate the effectiveness of ME and MDI constraints.
	\end{itemize}
	
	The rest of this paper is organized as follows. Section \ref{sec_related} introduces three types of closely related work, content-aware recommender systems in Section \ref{sec_related_con}, cross-domain recommendations in Section \ref{sec_related_cross} and meta-learning based recommender systems in Section \ref{sec_related_meta}. In Section \ref{sec_preli}, we introduce the problem formulation and notations in Section \ref{sec_preli_pro}, and briefly state the recommender systems equipped with the meta-learning framework in Section \ref{sec_preli_meta}. Next, we give a detailed introduction of the proposed three-block MetaDPA in Section \ref{sec_method}. Specifically, we firstly introduce the multi-source cross-domain adaptation in Section \ref{sec_method_multi} including MDI and ME constraints added on the domain adaption; we then introduce the diverse preference augmentation in Section \ref{sec_method_diverse} after multi-source domain adaptation; finally we introduce the preference meta-learning framework in Section \ref{sec_method_preference}. Then we demonstrate the experiments of the proposed MetaDPA together with the competing baselines in Section \ref{sec_exp}, which includes introductions of the experimental settings in Section \ref{sec_exp_set}, the overall experimental results compared with baselines in Section \ref{sec_exp_over}, the ablation studies to evaluate the effectiveness of MDI and ME constraints in Section \ref{sec_exp_abl}, and the impact of hyper-parameters $\beta_1$, $\beta_2$ the above two constraints in Section \ref{sec_exp_rob}. Finally, we conclude this paper in Section \ref{sec_con}.
	
	\section{Related Work}\label{sec_related}
	In this section, we focus on closely related work to our method: content-aware recommender systems, cross-domain recommender systems, and mete-learning based recommender systems. The proposed MetaDPA is a kind of cross-domain recommendation that transfers preference patterns from multiple source domains to a target domain based on a meta-learning optimization framework with content information.
	\subsection{Content-aware Recommender Systems} \label{sec_related_con}
	Content-aware recommender system, as a representative content-aware method, is widely studied by scholars recently. Such as the state-of-the-art collaborative deep learning (CDL)~\cite{wang2015collaborative} was proposed as a probabilistic model by jointly learning a probabilistic stacked denoising auto-encoder (SDAE)~\cite{vincent2010stacked} and CF. CDL exploits the interaction and content data to alleviate cold start and data sparsity problems. However, unlike CTR, CDL takes advantage of deep learning framework to learn effective real latent representations. Thus, it can also be applied in the cold-start and sparse settings. CDL is also a tightly coupled method for recommender systems by developing a hierarchical Bayesian model.
	
	Another classic content-aware method is deep cooperative neural networks (CoNN)\cite{zheng2017joint}, which consists of two parallel neural networks: one learns user behaviors exploiting users' reviews, and the other one learns  item properties from the items' reviews. A shared layer is introduced on the top to couple these two networks together. Then, dual attention mutual learning (DAML)\cite{liu2019daml} integrates ratings and reviews into a joint neural network with a local and mutual attention mechanism to strengthen the interpretability. In addition, higher-order nonlinear interaction of features are extracted by the neural factorization machines to predict ratings. 
	\subsection{Cross-domain Recommender Systems} \label{sec_related_cross}
	Cross-domain recommender systems, as a type of methods for solving cold-start and sparse challenges, transfer preference information from source domains to its related target domain and then improve the performance of recommendations in the target domain. These methods can be categorized into single source and multiple source cross-domain recommendations based on the number of source domains applied for transferring. 
	
Cross-domain recommendations with single source domain, such as cross-domain triadic factorization (CDTF)\cite{DBLP:conf/www/HuCXCGZ13}, deep domain adaptation model (DARec)\cite{DBLP:conf/ijcai/YuanYB19} , and equivalent transformation learner (ETL)\cite{DBLP:journals/corr/abs-2009-06884} were proposed to transfer user-item preference relations from a single source domain to a target domain without relying on any auxiliary information. By combining content information, a transfer meeting content-aware method (TMH)\cite{DBLP:conf/www/HuZY19} is formulated with unstructured text in an end-to-end manner.  Then, Cross-domain recommendation framework via aspect transfer network (CATN)\cite{zhao2020catn} is developed via an aspect transfer network for cold-start users. Another related study is text-enhanced domain adaptation recommendation (TDAR)\cite{DBLP:conf/kdd/YuLGOQ20} that extracts the textual features in word semantic space for each user and item and feeds them into a domain classifier with the embeddings of users and items for better domain adaptation.
	
For multi-source cross-domain recommendations, one pioneer work is Collective Matrix Factorization (CMF)\cite{singh2008relational} that extends linear models to arbitrary relational domains. Then, multi-domain collaborative filtering (MCF)\cite{DBLP:conf/uai/ZhangCY10} is proposed by considering multiple collaborative filtering tasks in different domains simultaneously and exploiting the relationships between domains. MCF also introduces the link function for different domains to correct their biases. Recently, MINDTL\cite{he2018robust} exploits transfer learning and compresses the knowledge from the source domain into a cluster-level rating matrix to model user's rating pattern. The rating patterns in the low level matrix are transferred to the target domain by enriching rating patterns by relaxing the full rating restriction on source domains. 
%	Then, Domain Switch-Aware Holistic Recurrent Neural Network (DS-HRNN)\cite{kim2019domain} models user sequential behavior across multiple domains and apply a single RNN-based framework to fully leverage multi-domain user behavior. A domain switch phenomena and DS-HRNN were proposed on top of the RNN-based framework by devising two domain switch-aware techniques to boost its predictability. 

%	\vspace{4pt}
	\subsection{Meta-learning Based Recommender Systems}\label{sec_related_meta}
%	\vspace{6pt}
Recently, many scholars focus on formulating recommendation frameworks with meta-learning\cite{lee2019melu, DBLP:conf/ijcnn/Bharadhwaj19, DBLP:conf/kdd/DongYYXZ20} for solving cold-start and sparse issues. Current meta-learning-based recommender systems adopt optimization-based meta-learning methods, such as model-agnostic meta-learning (MAML)\cite{DBLP:conf/icml/FinnAL17}, to improve the performance of new tasks. One representative work is MeLU\cite{lee2019melu} that utilizes MAML to rapidly adapts to new tasks with a few ratings. Another straightforward application of MAML on the recommender system is MetaCS\cite{bharadhwaj2019meta}, which formulates each user's rating pattern as a task in MAML and designs a recommendation framework based on this. Then, Memory-augmented meta-optimization (MAMO)\cite{DBLP:conf/kdd/DongYYXZ20} designs two memory matrices to store task-specific memories for parameter initialization and feature-specific memories for fast predicting users' preferences, respectively. By integrating heterogeneous information, MetaHIN\cite{DBLP:conf/kdd/Lu0S20} exploits meta-learning to alleviate the cold-start problem at both the data and model levels, which leverages multifaceted semantic contexts and a co-adaptation meta-learner to learn finer grained semantic priors for new tasks in both semantic and task-wise manners. By dynamic subgraph sampling to construct representative training tasks, MetaCF\cite{wei2020fast} dynamically samples subgraph centered at a user in the training phase to account for the effect of limited interactions under cold-start settings, and extends the historical interactions by incorporating potential interactions to avoid the overfitting problem. It can achieve good performance under cold-start settings. However, the drawback is obvious due to the introduction of potential interaction leading to low efficiency of training the model.

	As a promising approach, meta-learning frameworks can fast adapt to new users' or new items' recommendations. However, existing meta-learning based methods suffer poor performance caused by meta-overfitting on sparse interactive data. So in this paper we aim to solve the meta-overfitting problems in meta-learning based recommendations by the proposed diverse preference augmentation technique as introduced in Section \ref{sec_intro}.
	
	%as an effective tool to deal with the cold-start problems.  existing meta-learning-based recommendations, as discussed above, only consider directly constructing tasks from the historical data and ignore the meta-overfitting problem. In consequence, it may lead to poor performance in some cold-start scenarios (see MeLU in Section~\ref{sec:exp}, which shows worst performance of all baselines in almost all metrics, due to serious meta-overfitting). MetaCAR carefully considers the meta-overfitting problem by utilizing the meta-augmentation process and constructing mutually exclusive training tasks from historical and meta-augmentation data. It shows an outstanding potential performance advantage on other meta-learning-based methods, such as MeLU in our experiment.
%	\vspace{4pt}
	\section{Preliminaries}\label{sec_preli}
%	\vspace{4pt}
	In this section, we first describe the basic notations used in this paper and the problem formulation in Section \ref{sec_preli_pro}. Then we briefly introduce a general framework of meta-learning based recommender systems in Section \ref{sec_preli_meta}.
	\subsection{Problem Formulation} \label{sec_preli_pro}
%	\vspace{4pt}
  In this paper, we aim to provide recommendations in the target domain by transferring preference knowledge from source domains. We suppose there are $n$ users and $m$ items in the target domain, and their index sets are denoted as $U=\{1,\cdots ,n\}$ and $I=\{1,\cdots ,m\}$, respectively. The available data includes user-item interactive matrix $\mathbf{R}=\{r_{ui}\ge 0: u\in U, i\in I\}$, the user content data $\mathbf{C}^{U}$, and the item content data $\mathbf{C}^{I}$. If user $u$ has an interaction with item $i$, then $r_{ui}=1$; otherwise, $r_{ui}=0$. For each user $u$, the content data $\mathbf{c}_{u} \in \mathbf{C}^{U} $ is composed of bag-of-vectors generated from her/his rated items' review data. Similarly, the content data of each item $\mathbf{c}_{i} \in \mathbf{C}^{I}$ is extracted from all obtained reviews. 
%The main notations are listed in Table \ref{tab:notation}.

%	denotes the observed user-item interaction matrix (or preference pattern). The interactions could be explicit ratings, such as rating 1 to 5, and binary implicit feedback. As implicit feedback is more common, thus we design our e, because it's challenging to learn effective user/item representations with insufficient interactive data.model with binary implicit feedback,    The performances of recommendations only based on interactions are heavily dependent on the amount of observed interactions, and sparse interactions often leads to poor performanc
	
In this work, we divide users' set $U$ as $U_{e}$ and $U_{n}$, where we define $U_{e}$ as `\textit{existing users}', and each $u\in U_{e}$ represents the user who rates no less than 5 items, i.e., $|\{r_{ui}: u\in U_{e}\}|\ge 5$. The remaining users $u\in U_{n}$ are defined as `\textit{new users}' (or cold-start users). Similarly, we divide items' set $I$ as $I_{e}$ and $I_{n}$, where we define $ I_{e}$ as `\textit{existing item}s', and each $i\in I_{e}$ denotes the item of receiving no less than 5 ratings, i.e., $|\{r_{ui}: i\in I_{e}\}|\ge 5$, and the remaining items $i\in I_{n}$ are defined as `\textit{new items}' (or cold-start items). Then, we define the following four recommendation problems including the sparse issue (`Warm-start') and three kinds of cold-start issues solved in this paper:
		 \begin{itemize}
		 \item[(1)] \textbf{Warm-start}: how to improve the accuracy of recommending \textit{existing items} to \textit{existing users} with sparse interactions available? Given sparse ratings $R_{w}$  of users $U_{e}$ to items $I_{e}$, i.e., $R_{w}=\{r_{ui} > 0, u\in U_{e}, i\in I_{e}\}$, then we train our model $f(\theta): U_e \times I_e \rightarrow R_{w}$ on $R_{w}$. The goal of `Warm-start' is to predict unknown ratings $\hat{r}_{ui} \notin R_{w}$ of \textit{existing users} $U_{e}$ to \textit{existing items} $I_{e}$ with the trained model $f(\theta)$ and recommend $k$ \textit{existing items} with top-$k$ ratings to each user in $U_{e}$. 
		\item[(2)] \textbf{C-U}: how to improve the accuracy of recommending \textit{existing items} to \textit{new users}? After training our model $f(\theta)$ with $R_{w}$, we finetune $f(\theta)$ with only a few ratings $R_{cu}$ of \textit{new users} $u\in U_{n}$ to \textit{existing items} $i\in I_{e}$. The goal of `C-U' is to predict unknown ratings $\hat{r}_{ui}\notin R_{cu}$ of \textit{new users} $U_{n}$ to \textit{existing items} $I_{e}$ by the finetuned model $f(\theta)$ and recommend $k$ \textit{existing items} with top-$k$ ratings to \textit{new users}.
		\item[(3)] \textbf{C-I}: how to improve the accuracy of recommending \textit{new items} to \textit{existing users}? Similar to \textbf{C-U}, we finetune $f(\theta)$ with only a few ratings $R_{ci}$ of \textit{existing users} $U_{e}$ to \textit{new items} $ I_{n}$. The goal of `C-I' is to predict unknown ratings $\hat{r}_{ui}\notin R_{ci} $ of \textit{existing users} $U_{e}$ to \textit{new items} $I_{n}$ by the finetuned model $f(\theta)$ and recommend $k$ \textit{new items} with top-$k$ ratings to \textit{existing users}.
		\item [(4)] \textbf{C-UI}: how to improve the accuracy of recommending \textit{new items} to \textit{new users}? We finetune $f(\theta)$ with a few ratings $R_{cui}$ of \textit{new users} $U_{n}$ to \textit{new items} $ I_{n}$. The goal of `C-UI' is to predict unknown ratings $\hat{r}_{ui}\notin R_{cui}$ of \textit{new users} $U_{n}$ to \textit{new items} $I_{n}$ with the finetuned model and recommend $k$ \textit{new items} with top-$k$ ratings to \textit{new users}.
	\end{itemize}

%	This paper focuses on the cold-start recommendations in the target domain. Given content data $\mathbf{c}_{u} \in \mathbf{C}^{U}$ of a user $u$ and $\mathbf{c}_{i} \in \mathbf{C}^{I}$ of an item $i$, and sparse rating matrix $\mathbf{R}_t$. Users or items with less than 5 interactions are treated as the cold-start users or cold-start items. The recommendation tasks on these cold-start users or items are called user/item cold-start recommendations. The goal of this paper is to conduct the following three types of cold-start recommendations in the target domain by transferring preference properties from multiple source domains: (1) recommending unseen items in the training stage to existing users (`C-I'); (2) recommending existing items to unseen users (`C-U'); (3) recommending unseen items to unseen users (`C-UI'). 
			\begin{figure*}[!htp]
	\centering
	\includegraphics[width=0.9\linewidth]{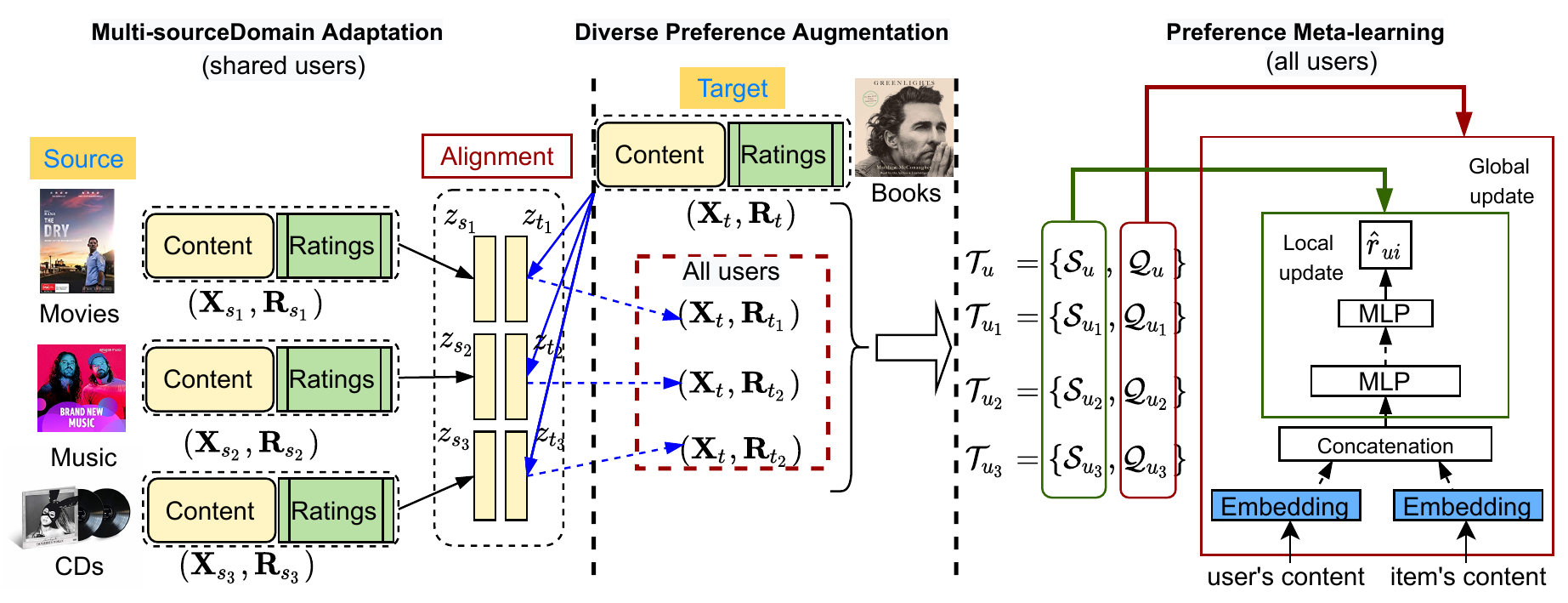}
	\vspace{-4pt}
	\caption{The framework of diverse preference augmentation with multiple source domains based on meta-learning (MetaDPA) consists of three blocks: Multi-source Domain Adaptation, Diverse Preference Augmentation, and Preference Meta-learning. We first train the framework of multi-source domain adaptation; and then we generate diverse ratings for the target domain given the content data; and finally we learn the preference model formulated as a multi-layer architecture based on the meta-learning optimization framework MAML. }
			\vspace{-8pt}
	\label{fig:framework}
\end{figure*}		
%	\vspace{4pt}
	\subsection{Meta-learning for Recommendations} \label{sec_preli_meta}
%	\vspace{4pt}
	The objective of meta-learning is to learn good initial weights for a model that can fast adapt to previously unseen tasks with a few samples\cite{DBLP:conf/nips/VartakTMBL17}. In meta-learning based recommendations\cite{lee2019melu,lu2020meta}, a user's preference prediction over all items is treated as a meta-learning task $\mathcal{T}_{u}$, which is denoted as the dataset used for the task. To be specific, $\mathcal{T}_{u}=(\mathbf{c}_{u}, \mathbf{r}_{u})$, where $\mathbf{c}_{u}$ is the input of all items for a specific user $u$ and $\mathbf{r}_{u}$ is the ratings of a user to all items. The input identifies a user, and it could be her/his identity, profile, historical preference information, or the review comments written by the user. In this paper, we user the review comments as the input. In particular, according to the setting of meta-learning, we divide  into a support set $\mathcal{S}_{u}$ and a query set $\mathcal{Q}_{u}$. Tasks from all users' preferences are randomly splited  into two disjoint partitions: one for meta-training tasks denoted as $\mathbb{T}_{tr}$, and another one for meta-testing tasks denoted as $\mathbb{T}_{te}$. The meta-learning based recommendations aim to learn a good preference prior distribution on meta-training tasks, and then can fast adapt to new tasks in meta-testing tasks.

	Suppose a supervised learning problem considers training on a dataset for a single task $\mathcal{T}$. In contrast, meta-learning considers to learn a set of tasks ${\mathcal{T}_u} \in \mathbb{T}_{tr}$, which are sampled from the task distribution $p(\mathcal{T})$. The objective of meta-training for recommendations is formulated as follows:
	\begin{align}
	\mathop{min}_{\theta}\sum_{\mathcal{T}_u\sim \mathbb{T}_{tr}}\mathcal{L}_{\mathcal{T}_u}(\mathcal{\theta} - \alpha \nabla_{\theta} \mathcal{L}_{\mathcal{T}_u}(\mathcal{\theta} , \mathcal{S}_{u}), \mathcal{Q}_{u}),
	\end{align}
	where $\nabla_\theta$ denotes the gradient w.r.t parameters $\theta$ of a preference prediction model, and $\alpha$ is the meta-learning rate, and  $\mathcal{\theta} - \alpha \nabla_{\theta} \mathcal{L}_{\mathcal{T}_u}(\mathcal{\theta} , \mathcal{S}_{u}$ is the task-specific parameters adapted to the task $\mathcal{T}_{u}$ after one gradient step from the global $\theta$. The algorithm locally updates parameters $\theta$ based on the gradient with $\mathcal{S}_u$, and then globally updates $\theta$ based on $\mathcal{Q}_{u}$, so that the globally updated parameters adapt to various tasks\cite{lee2019melu}. During meta-testing, the meta-learner adapts the learned $\theta$ to $\mathcal{T}_{u} \in \mathbb{T}_{te}$ with its support set $\mathcal{S}_u$, and then the adapted parameters $\theta$ is used to predict ratings and evaluate the recommendation performance in its query set $\mathcal{Q}_u$\cite{lu2020meta}.
%\vspace{4pt}
	\section{Methodology}\label{sec_method}
%	\vspace{4pt}
	In this section, we propose a novel method Diverse Preference Augmentation with multiple domains (MetaDPA) for cold-start recommendations as illustrated in Fig.~\ref{fig:framework}. Our method consists of three blocks: \textit{Multi-source Domain Adaptation}, \textit{Diverse Preference Augmentation}, and \textit{Preference Meta-learning}.
	
	Firstly, we conduct multi-source domain adaptation between the source domains and the target domain with Dual CVAEs, and align the representations of the source and target domains by the principle of InfoMax\cite{federici2020learning} in Section \ref{sec_method_multi}. Then, we employ the learned encoder-decoders to generate diverse ratings, which is named as diverse preference augmentation in Section \ref{sec_method_diverse}. Next, these augmented diverse ratings together with the original true ratings are fed into the preference meta-learning procedure in Section \ref{sec_method_preference}. Finally, we test the recommendation performances under three cold-start settings separately.

%	\vspace{4pt}
	\subsection{Multi-source Domain Adaptation}\label{sec_method_multi}
%	\vspace{4pt}
	In this section, our objective is to conduct domain adaptation from multiple source domains to the target domains. As users'/items' content information is commonly used for alleviating cold-start issues, and the content data including items' images, descriptions and users' profile, user-item review texts. Most of the content data are not shared by different domains. In TDAR\cite{DBLP:conf/kdd/YuLGOQ20}, it takes the review texts as domain-invariant features to align the latent space for domain adaptation. To facilitate the domain adaptation, we also adopt the domain-invariant reviews as the content data. Specifically, the user content $\mathbf{c}_{u}$ is the collection of reviews from items rated by the user $u$. We first encode the user's content into a dense low-dimensional embeddings $\mathbf{x}_{u}$. We denote $\mathbf{x}_{u}^{(s)}$ and $\mathbf{x}_{u}^{(t)}$ as the user content vectors in the source domain and in the target domain, respectively. For simplicity, we use $\mathbf{x}_{s}$ and $\mathbf{x}_{t}$ to represent them. Similarly, we denote $\mathbf{r}_{s}$ and $\mathbf{r}_{t}$ as ratings rated by the shared user $u$ to items in the source domain and the target domain, respectively.
	
	We use a Dual-CVAE network to learn users' latent representations and reconstruct ratings $\mathbf{r}_{s}$ and $\mathbf{r}_{t}$. In the Dual-CVAE, we add conditions $\mathbf{x}_s$ ($\mathbf{x}_t$) on the latent representations in a similar way with HCVAE\cite{wu2020hybrid}. Firstly, the input $\mathbf{x}_s$ and $\mathbf{x}_t$ are encoded into a distribution $q_{\phi_s}(\mathbf{z}_s|\mathbf{r}_{s},\mathbf{x}_{s})$ and $q_{\phi_t}(\mathbf{z}_t|\mathbf{r}_{t},\mathbf{x}_{t})$, respectively. The encoders of the Dual-CVAE maps ratings $\mathbf{r}_s$ and $\mathbf{r}_t$ to latent representations $\mathbf{z}_s$ and $\mathbf{z}_t$, respectively. The optimization objective of the Dual-CVAE is the evidence lower bound (ELOB)\cite{doersch2016tutorial}, which consists of the sum of the reconstruction error, namely, maximizing the likelihood estimation of the decoders $\log p_{\theta_s}(\mathbf{r}_s|\mathbf{z}_s, \mathbf{x}_s)$ and $\log p_{\theta_t}(\mathbf{r}_t|\mathbf{z}_t, \mathbf{x}_t)$, and the negative KL-divergence between the variational posterior and the prior. So the loss function in the source domain and the target domain can be written as follows\cite{wu2020hybrid}:
	\begin{align} \label{elbo}
	\mathcal{L}_{\textit{ELBO}} &= \mathcal{L}(\mathbf{r}_s, \mathbf{x}_s; \theta_s, \phi_s) + \mathcal{L}(r_t, \mathbf{x}_t; \theta_t, \phi_t) \nonumber\\
	&= \mathbb{E}_{q_{\phi_s}(\mathbf{z}_s|\mathbf{r}_s, \mathbf{x}_s)}[\log p_{\theta_s}(\mathbf{r}_{s}|\mathbf{z}_s, \mathbf{x}_s)] \nonumber\\
&	+\mathbb{E}_{q_{\phi_t}(\mathbf{z}_t|\mathbf{r}_t, \mathbf{x}_t)}[\log p_{\theta_t}(\mathbf{r}_{t}|\mathbf{z}_t, \mathbf{x}_t)]  \nonumber \\
	&- D_{KL}[q_{\phi_s}(\mathbf{z}_s|\mathbf{r}_s, \mathbf{x}_s)||p(\mathbf{z}_s)] \nonumber \\
	& -D_{KL}[q_{\phi_t}(\mathbf{z}_t|\mathbf{r}_t, \mathbf{x}_t)||p(\mathbf{z}_t)].
	\end{align}
We employ the softmax function as the activation function in the output layer that maps the reconstructed output into the range of [0,1], which is consistent with the range of reconstructed ratings. The reconstruction loss between the predictions and true ratings could be MSE if the user-item interactions are explicit feedback and binary cross entropy if the
user-item interactions are implicit feedbacks\cite{chen2009collaborative}. In this paper, we use the implicit feedback as ratings, so we adopt the binary cross-entropy as the reconstruction loss.

The KL divergence loss of the Dual-CVAE can be estimated using the Stochastic Gradient Variational Bayes (SGVB) estimator\cite{kingma2013auto}. Besides, we hope to learn latent representations from the content embeddings $\mathbf{x}_s$ and $\mathbf{x}_t$, and so the learned distributions of latent representations store preference information from both ratings and content, which makes it capable of reconstructing ratings only using content. So we have the following objective: 
	\begin{align} \label{kl}
	\mathcal{L}_{\textit{KL}}=&-\frac{1}{2}\sum_{l=1}^{L}[\mathbf{\sigma}_{s_l}^2+(\mathbf{\mu}_{s_l}-\mathbf{z}_{s_l}^x)^2 - {\rm \log}\mathbf{\sigma}_{s_l}^2-1] \nonumber \\
&	+\frac{1}{2}\sum_{l=1}^{L}[\mathbf{\sigma}_{t_l}^2+(\mathbf{\mu}_{t_l}-\mathbf{z}_{t_l}^x)^2 - {\rm \log}\mathbf{\sigma}_{t_l}^2-1].
	\end{align}
	where $L$ is the dimension of the latent sampled representations. For the source domain, $ \mathbf{u}_s$ and $ \mathbf{\sigma}_s$ are the mean and variance of the approximate posterior, and $\mathbf{z}_s^x$ is the output of a dense embedding encoder $E_s^x$, parameterized by $\phi_{x_s}$. The notations $ \mathbf{u}_t$, $ \mathbf{\sigma}_t$, $\mathbf{z}_t^x$, $E_t^x$, and $\phi_{x_t}$ in the target domain have similar meanings.

To reconstruct ratings only using content, we align the latent representations of the Dual-CVAE to the output dense embedding vector of encoder $E_s^x$ by the following mean square error (MSE) loss:
	\begin{equation} \label{mse}
	\mathcal{L}_{\textit{MSE}}
	=||\mathbf{z}_s-q_{\phi_{x_s}}(\mathbf{z}_s^x|\mathbf{x}_s)||^2+||\mathbf{z}_t-q_{\phi_{x_t}}(\mathbf{x}_t^x|\mathbf{x}_t)||^2.
	\end{equation}

For domain adaptation, we construct the cross-domain reconstruction objectives by generating ratings $\mathbf{r}_s$ of the source domain with the latent representation of the target domain $\mathbf{z}_t$, and reconstructing ratings $\mathbf{r}_t$ of the target domain with the latent factors of the source domain $\mathbf{z}_s$. To align the latent representations $\mathbf{z}_s$ and $\mathbf{z}_t$, we have the following cross-domain reconstruction loss,
	\begin{align}\label{sim}
	\mathcal{L}_{\textit{Rec}} = &\frac{1}{2m}\sum_{i=1}^{m}[r_{si}\log(\hat{r}_{ti})+(1-r_{si})\log (1-\hat{r}_{ti})] \nonumber\\
	& + \frac{1}{2m}\sum_{i=1}^{m}[r_{ti}\log(\hat{r}_{si})+(1-r_{ti})\log (1-\hat{r}_{si})],  
	\end{align}
where $\hat{r}_{ti}$ is the reconstructed rating of user $u$ to item $i$ by feeding the latent representation of target domain $\mathbf{z}_t$ into the decoder $D_s$ in the target domain as shown in the Fig.~\ref{fig:generation_network}, and $\hat{r}_{si}$ is the generated rating of user $u$ to item $i$ by injecting the latent representation of source domain $\mathbf{z}_s$ into the decoder of target domain $D_t$.

To maximize the mutual information from the source and target domains and preserve the domain-specific properties, we impose the MDI constraint with the InfoMax principle\cite{hjelm2018learning} on sampled latent representations $\mathbf{z}_s$ and $\mathbf{z}_t$. Since the MDI constraint does not discard superfluous information of each domain for predicting ratings. So it allows the reconstructed user's preference to have the capacity of preserving the domain-specific properties as well as the domain-shared properties. The MDI constraint is written as follows:
	\begin{equation}\label{mdi}
	\mathcal{L}_{\textit{MDI}}(\phi_s,\phi_t) =-I_{\phi_s,\phi_t}(\mathbf{z}_s,\mathbf{z}_t),
	\end{equation}
where the function $I(\cdot)$ denotes the mutual information between inputs\cite{federici2020learning}. The MDI is employed here because we consider each user’s latent preferences $\mathbf{z}_s$ and $\mathbf{z}_t$ encoded from different domains should have both domain-shared and domain-specific properties. So the MDI advocates the decoders of each domains to generate domain-specific ratings, which is consistent with the goal of generating diverse ratings by the content in the target domain.
%	\vspace{4pt}
	\subsection{Diverse Preference Augmentation}\label{sec_method_diverse}
	As introduced in Section \ref{sec_intro}, the goal of diverse preference augmentation is to generate diverse ratings by the learned encoder-decoders (highlighted with red line shown in Fig.~\ref{fig:generation_network}) of target domain given the input $\mathbf{x}_t$. If there are $k$ source domains, we have $k$ Dual-CVAEs for the multi-source domain adaptation, then we generate $k$ ratings for each user in the target domain by his/her content $\mathbf{x}_t$. Due to the integrated effect of MDI constraint imposed on the latent representations, the reconstruction loss \eqref{elbo} and the cross-domain reconstruction loss \eqref{sim}, the Dual-CVAE can learn domain-shared preference information well enough. Thus, these generated $k$ ratings from $k$ Dual-CVAEs are similar because all source domains adapts to the target domain. To avoid overfitting, we impose the ME constraints on the outputs of $k$ decoders $D_t$ to generate ratings as different as possible from each other. With the ME constraint, we force the generated rating from the decoder $D_t$ as close to the reconstructed ratings as possible from the decoder $D_s$. Thus, the generated ratings preserve domain-specific preference patterns of multiple source domains, so we can generate diverse ratings by $k$ encoder-decoder frameworks of Dual-CVAEs. The ME constraint is realized by maximizing the mutual information between two generated ratings $\mathbf{r}_s$ and $\mathbf{r}_t$ of a shared user $u$ as follows:
	\begin{equation}\label{me}
	\mathcal{L}_{\textit{ME}} = -I(\mathbf{r}_s, \mathbf{r}_t),
	\end{equation}
	where $I(\cdot)$ denotes the mutual information between two inputs, that is implemented by InfoNCE\cite{oord2018representation}. 
	
    By enforcing ME and MDI constraints on the Dual-CVAE, the objective of cross-domain adaptation can be derived by summing up objectives \eqref{elbo}, \eqref{mse}, \eqref{sim}, and two constraints \eqref{mdi}, \eqref{me} together. Single-source cross-domain adaptation is one special case of multi-source cross-domain adaptation. We can obtain the multi-source cross-domain objective by integrating the following cross-domain objective together,
\begin{equation} \label{dual}
\mathcal{L}_{\textit{Dual-CVAE}} =  \mathcal{L}_{\textit{ELBO}} + \mathcal{L}_{\textit{MSE}} + \mathcal{L}_{\textit{Rec}} + \beta_1\mathcal{L}_{\textit{MDI}}+ \beta_2\mathcal{L}_{\textit{ME}}.
\end{equation}

We train the Dual-CVAE shown in Fig.~\ref{fig:generation_network} by minimizing the objective \eqref{dual} of cross-domain adaptation. The multi-source cross-domain adaptation can be implemented by training multiple Dual-CVAEs in parallel. Suppose we have $k$ source domains, then we learn $k$ Dual-CVAEs independently. After that, we obtain the learned $k$ encoders $E_{t}^{x}$ and $k$ decoders $D_t$, and then employ them to generate $k$ diverse ratings $\mathbf{r}_{t1}, \mathbf{r}_{t2}, \cdots, \mathbf{r}_{tk}$ with the content $\mathbf{x}_{t}$. It's worth note that these generated ratings are in the continuous scale of $[0, 1]$ because we focus on implicit feedback in this paper.

By taking users' preferences / ratings prediction as meta-learning tasks, the meta-learner trained on these generated ratings and original ratings is expected to avoid overfitting as introduced in Section \ref{sec_intro}. In Section \ref{sec_related}, we know that the meta-learning optimization scheme MAML\cite{DBLP:conf/icml/FinnAL17} can fast adapt to new tasks. In recommendations, the meta-learner can fast adapt to cold-start recommendations with MAML.

%be fast adaptable to predict ratings of new users/items, so we introduce them into the preference meta-learning stage in the target domain to improve the performance of cold-start recommendations.
	\subsection{Preference Meta-learning}\label{sec_method_preference}
%	\vspace{4pt}
This block focuses on training a preference prediction model based on meta-learning in the target domain by generated diverse ratings $\mathbf{r}_{t1}, \mathbf{r}_{t2}, \cdots, \mathbf{r}_{tk}$. Firstly, we encode the original content $\mathbf{c}_u$ and $\mathbf{c}_i$ to their dense embedding $\mathbf{x}_{u}$ and $\mathbf{x}_{i}$. We denote $\mathbf{c}_t$ as the combination of $\mathbf{c}_u$ and $\mathbf{c}_i$ in the following paper. Then, we construct the meta-learning task $\mathcal{T}_u$ and the augmented tasks {$\mathcal{T}_{u}^1, \cdots , \mathcal{T}_{u}^k$} as follows:
	\begin{align}
	\mathcal{T}_u & ={(\mathbf{c}_t, \mathbf{r}_t)}
	\\
	\mathcal{T}_{u_1}&= {(\mathbf{c}_t,
		\hat{\mathbf{r}}_{t1})}, \cdots,
	\mathcal{T}_{u_k}= {(\mathbf{c}_t,
		\hat{\mathbf{r}}_{tk})},
	\end{align}
	where $\mathbf{r}_t$ is the original ratings in the target domain. Next, we learn the preference prediction model via the meta-learning framework MAML. 
	
	As shown in Fig.~\ref{fig:framework}, we firstly employ a fully connected embedding layer to encode content vectors $\mathbf{c}_u$ and $\mathbf{c}_i$ into dense embeddings $\mathbf{x}_u$ and $\mathbf{x}_i$, and then we adopt a multi-layer neural architecture\cite{he2017neural2} to predict rating scores by the concatenation of $\mathbf{x}_u$ and $\mathbf{x}_i$. As implicit feedback we considered in this paper, so we use the binary cross-entropy loss on the top of multi-layer neural network to predict ratings.

%	 The embedding layer and the multi-layer architecture forms the preference prediction framework After introducing augmented ratings and original true ratings into the target domain, the rating $r_{ui}$ can be from original ratings or augmented ratings. 
The embedding layer and the multi-layer architecture constitute the preference prediction model. We train the model with diverse tasks to avoid overfitting via the meta-learning framework MAML. The objective can be written as:
	\begin{equation}
	\hat{\mathbf{r}}_{ui} = f(\theta_l, \theta_e, \mathbf{c}_{u}, \mathbf{c}_{i}),
	\end{equation}
	where $\theta_e$ is the parameters of the fully connected embedding layer that encodes the content vector $\mathbf{c}_u$ and $\mathbf{c}_i$ into dense embeddings $\mathbf{x}_u$ and $\mathbf{x}_i$, and $\theta_l$ is the parameters of the multi-layer neural network. 
		
%	The MAML can learn the preference distribution of all users in the target domain. As implicit feedback is taken into consideration in our work, 
	
	As shown in Fig. ~\ref{fig:framework}, MAML includes the inner loop for local update of the model and the outer loop for global update of the model. In this work, we train the MAML on training task set $\mathbb{T}_{tr}$, which includes the original tasks $\mathcal{T}_{u}$ and augmented tasks $\mathcal{T}_{u1}, \cdots, \mathcal{T}_{uk}$. We divide samples of each task $\mathcal{T} \in \mathbb{T}_{tr}$ randomly into a support set $\mathcal{S}_u$ and a query set $\mathbf{Q}_u$ as follows:
	\begin{align}
	\mathcal{T} = \{\mathcal{S}_{u}, \mathcal{Q}_{u}\}.
	\end{align}
%	We select users or items with no less than 5 original true ratings as active users for meta-training, and the remaining users are treated as cold-start users or items for meta-testing. 
	
	In the meta-testing phase, meta-testing tasks $\mathcal{T} \in \mathbb{T}_{te}$ includes only original tasks constituted from original ratings of the target domain. Samples of each $\mathcal{T}$ are also split into a support set $\mathcal{S}_u$ for fine-tuning the preference model and a query set $\mathbf{Q}_u$ for testing the recommendation performance. 
%	We use metrics for evaluating top-k recommendations and report the experimental results in Section~\ref{sec_exp}.

\subsection{Time Complexity Analysis} \label{sec-time}
Our model consists of three blocks (Fig.\ref{fig:framework}). (1) the Dual-CVAE (Fig.~\ref{fig:generation_network}) is trained in parallel for multiple source domains. It encodes users’ ratings and users’ content into latent vectors $\mathbf{z}_{s}, \mathbf{z}_{t}$ with 2-layer networks and then decodes the latent vectors into users’ ratings with 2-layer networks. The dimensions of users' content, hidden layers and the latent vectors are constant with the data size $\mathcal{O}\{\text{max}\{n, m\}\}$, where $n$ and $m$ are numbers of users and items. If we denote $B$, $l$ and $m$ as the batch size, the numbers of items in the source and target domains, then the dimensions of inputs $\mathbf{r}_s$ and $\mathbf{r}_t$ are $l$ and $m$, respectively. So the time complexity is $\mathcal{O}(B(l+m))$. (2) The second block forwards the encoder and decoder (red line in Fig.~\ref{fig:generation_network}) one-pass to generate ratings, so the time complexity is $\mathcal{O}(B)$. (3) The third block learns a 2-layer network, and the dimension of users' (items') content is constant with the data size, so the time complexity is $\mathcal{O}(B)$. Overall, the time complexity is $\mathcal{O}(B(l+m))$, so it scales linearly with the data size.

	\begin{table}
		\caption{Statistics of datasets for multiple source domains.}
\vspace{-8pt}
		\centering
		\begin{tabular}{c|c|c|c|c|c}
			\toprule
			\multirow{2}{*}{Source ($S$)} & \multicolumn{2}{c|}{\#shared users ($T$)} & \multirow{2}{*}{\#items} & \multirow{2}{*}{\#ratings} & \multirow{2}{*}{sparsity} \\ \cline{2-3}
			& Books & CDs & & & \\ \hline
			Electronics & 28,505 & 6,260 & 63,001 & 1,687,993 & 99.98\% \\ \hline
			Movies & 37,387 & 18,031 & 50,052 & 1,697,438 & 99.97\% \\ \hline
			Music & 1,952 & 5,331 & 3,568 & 64,705 & 99.67\% \\ 
			\toprule
		\end{tabular}
	\vspace{-10pt}
		\label{tab:source}
	\end{table}
	
	\begin{table}%[!t]
		%\extrarowheight=3pt
		
		\caption{Statistics of datasets for target domains.}
		% 		\small
		\vspace{-8pt}
		\centering
		%\begin{tabular*}{0.41\paperwidth}{@{\extracolsep{\fill}}ccccccc}
		\begin{tabular}{c|c|c|c|c}
			\toprule
			Datasets& \#users&\#items&\#ratings & sparsity \\
			\hline
			Books         & 603,374   &348,957  & 8,575,000  & 99.99\% \\
			\hline
			CDs         & 25,400  & 24,904 & 43,903    & 99.99\% \\
			\toprule
		\end{tabular}
		\vspace{-10pt}
		\label{tab:target}
	\end{table}
	
	\section{Experiments}\label{sec_exp}
%	\vspace{4pt}
%    Existing studies for solving cold-start or sparse issues in recommendations mainly include three types: (1) content-aware recommendations that incorporate side information into the model, (2) cross-domain recommendations with transfer learning, and (3) meta-learning based recommender systems. 
    
    In this work, we claim that (1) diverse preference augmentation can handle overfitting in the case of sparse interactions to improve the performance of four recommendation problems defined in Section \ref{sec_preli_pro}; (2) learning a preference model via a meta-learning scheme can alleviate cold-start issues. The goal of experiments is to evaluate the effectiveness of the proposed MetaDPA to avoid overfitting and solve the recommendation problems including `Warm-start', `C-U', `C-I', and `C-UI'. In addition, the experiments study the impact of two constraints added in the model. To be specific, we analyze the effectiveness from the following aspects:
    \begin{itemize}
    	\item \textbf{RQ1:} Does our method outperforms the state-of-the-art cross-domain baselines?
    	\item \textbf{RQ2:} Does our method handle overfitting on sparse interactions?
    	\item \textbf{RQ3:} How is the scalability of our method?
    	\item \textbf{RQ4:} How do the two constraints MDI and ME affect the performances of recommendations under different settings?
    	\item \textbf{RQ5:} How do the hyper-parameters, such as $\beta_{1}$ of MDI, and $\beta_{2}$ of ME, affect the effectiveness of DPA?
    	%   Is meta-augmentation with the auxiliary domain truly effective for recommendations under cold-start circumstances?
    \end{itemize}

\begin{figure*}[!htbp]
	\centering
	\includegraphics[width=1\linewidth]{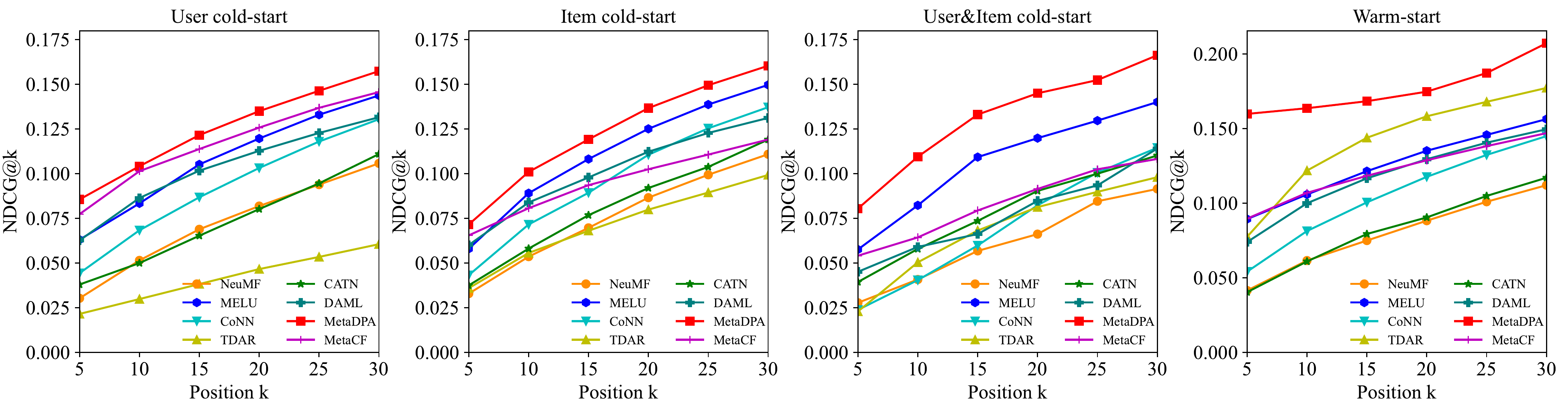}
	\vspace{-16pt}
	\caption{Performance comparison on Books.}
	% 		\small
		\vspace{-12pt}
	\label{fig_books}
\end{figure*}
\begin{figure*}[!htbp]
	\centering
	\includegraphics[width=1\linewidth]{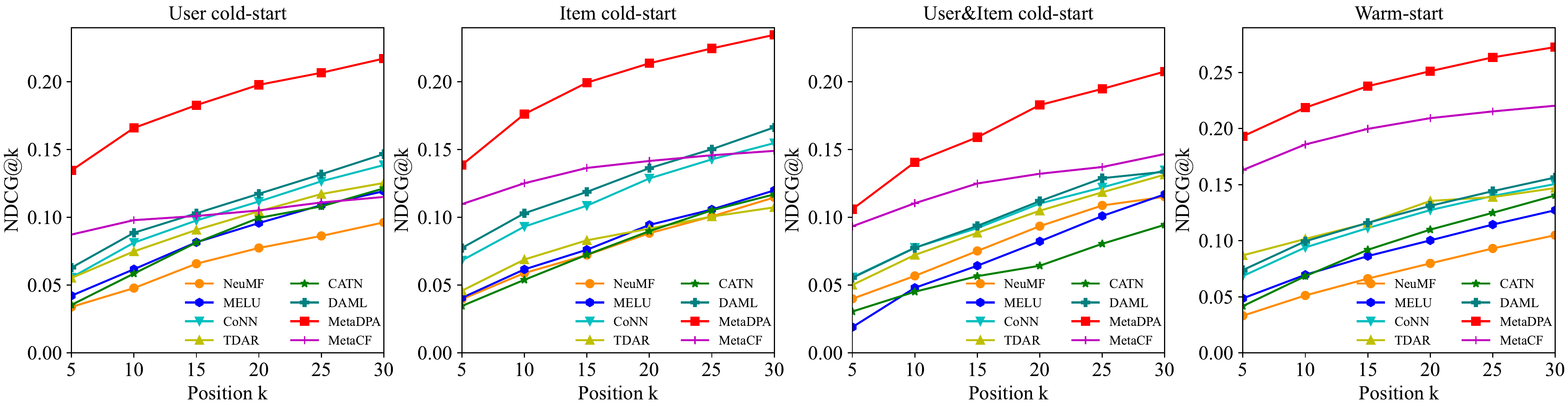}
		\vspace{-16pt}
	\caption{Performance comparisons on CDs.}
		\vspace{-12pt}
	\label{fig_cds}
\end{figure*}
%\vspace{4pt}
	\subsection{Experimental Settings} \label{sec_exp_set}
%	\vspace{4pt}
	\subsubsection{Datasets}
	We evaluate the performance of our method against the state-of-the-art baselines on public Amazon datasets\footnote{http://jmcauley.ucsd.edu/data/amazon/}. Among the largest categories, we choose Electronics, Movies and Music as three source domains, and Books and CDs as two target domains. As this paper focuses provide recommendations based on implicit feedbacks, we transform explicit ratings greater than 0 as the positive feedbacks '1' and others as negative feedbacks '0'. Statistics for the source domains and the target domains are shown in Table~\ref{tab:source} and Table~\ref{tab:target}, respectively. 
	
	Experiments of our method contains three phases: (1) multi-source domain adaptation from three source to two target domains, (2) diverse preference augmentation, and (2) preference meta-learning in the two target domains. In the first phase, we train three independent Dual-CVAEs of multi-source domain adaptation. In this phase, we discard users or items with fewer than 20 positive ratings for both source and target domains. We randomly split ratings into 80\% as training set and the remaining 20\% as evaluation set for the domain adaptation. Then, we generate diverse ratings from three decoders of Dual-CVAEs for users in the target domains by their content data. In the third phase, we train preference model with the generated ratings and the original ratings in the target domains, and test the performances of four problems defined in Section \ref{sec_preli_pro}. 
	
%	For the original ratings in the target domains, we divide items and users into two groups (existing/new) as Melu\cite{lee2019melu} to evaluate the performance under both cold-start and warm-start settings. Specifically, we define existing users/items that have no less than 5 positive ratings. The remaining users/items are defined as new users/items. We category four situations for the two target domains respectively: (1) existing items for new users, (2) new items for existing users, (3) new items for new users, and (4)) existing items for existing users. The meta-training tasks are created by the generated ratings and original ratings from (4). The meta-testing tasks of each situation are created by original ratings in each. evaluate the performances on different cold-start settings.	
%	\vspace{4pt}
	\subsubsection{Evaluation Protocols}
%	\vspace{4pt}
%	In this work, we focus on addressing three cold-start and traditional warm-start scenarios and in the two target domains:
%	 \begin{itemize}
%		\item \textbf{C-U}, refers to the cold-start user recommendation that recommends existing items to an unseen user in the meta-training step;
%		\item \textbf{C-I}, refers to the cold-start item recommendation that recommends an unseen item to existing users; 
%		\item  \textbf{C-UI}, refers to recommendation for unseen users with unseen items in meta-training block;
%		 \item \textbf{Warm-start}, refers to recommendations for existing users with existing items.
%	\end{itemize}  
	
In this paper, we only test performances on target domains. Recommendation problems consist of `C-U', `C-I', `C-UI', and `Warm-start' introduced in Section \ref{sec_preli_pro}. We first train our model on training tasks $\mathcal{T}_{tr}$ constructed of ratings in $\mathbf{R}_{w}$. For solving `Warm-start', we test the performance on the query set of $\mathcal{T}_{tr}$. For cold-start problems `C-U', `C-I', and `C-UI', we first finetune the trained model on training tasks $\mathcal{T}_{tr}$ by the support set of testing tasks $\mathcal{T}_{te}$, and then we test the performance on the query set of $\mathcal{T}_{te}$. Taking `C-U' as an example, $\mathcal{T}_{te}$ is constructed from ratings $\{r_{ui}: u\in U_n, i\in I_{e}\}$. Similarly, we can construct testing tasks $\mathcal{T}_{te}$ for other two cold-start settings.
%	in cold-start scenarios are sampled from the dataset for meta-testing. For each task in cold-start scenarios, we use 5 samples to fine-tune the model. To evaluate the performance of the proposed method, we use the unseen items from both meta-testing and meta-training tasks to report the experiment results.
	
Following the common strategy of implicit feedbacks\cite{he2017neural2}, we adopt the similar \textit{leave-one-out} evaluation protocols for evaluating the performance of recommendations. Specifically, we sample 99 negative unobserved items for each positive item to evaluate MetaDPA and other baselines. Then we report the performance of top-k recommendations under four common metrics: \textit{hit ratio} (HR)\cite{yin2013lcars}, \textit{mean reciprocal rank} (MRR)\cite{zhang2018discrete}, \textit{normalized discounted cumulative gain} (NDCG)\cite{lee2019melu}, and \textit{area under ROC curve} (AUC)\cite{rendle2009bpr}. 
%\vspace{4pt}
	\subsubsection{Baselines}
	
\begin{table*}[!htbp] \label{performance}
	%\extrarowheight=3pt
	\caption{The overall performance comparison of MetaDPA (the proposed model) and baselines. Best performance marked as bold font, the second best marked as $^{o}$.} 
	\vspace{-8pt}
	\centering
	\begin{tabular}{c|c|cccc||c ccc }
		%\hline
		%		\toprule[1pt]
		%\textbf{}& \lq Cloth\rq & & \lq Cell\rq & \\
		%		  \hline
		\toprule
		%		   \hlinefill
		\multicolumn{2}{c|}	{Dataset}&\multicolumn{4}{c||}{Books}&\multicolumn{4}{c}{CDs}\\
		\hline 
		\multicolumn{2}{c|}	{Metric} & HR@10&MRR@10 & NDCG@10 & AUC & HR@10&MRR@10& NDCG@10 & AUC\\
		\hline \hline
		
		\multirow{7}{*}{C-U}  & 
		NeuMF & 0.1159 & 0.0327 &  0.0515 & 0.5009 & 0.1069 &   0.0300  &  0.0478 & 0.5054  \\
		\cline{2-10}
		&MeLU & 0.1524 & 0.0630 &  0.0834 & \textbf{0.5819}& 0.1278 &  0.042  &  0.0617 & 0.5635 \\
		\cline{2-10}
		&CoNN & 0.1515 & 0.0437 &  0.0683 & 0.5810 & 0.1696 &  0.0549 &  0.0813 & 0.5853 \\
		\cline{2-10}
		&	TDAR & 0.0641 & 0.0195 &  0.0298 & 0.3852 & 0.1417 &  0.0549 &  0.0748 & 0.5451 \\
		\cline{2-10}
		&	CATN & 0.1071 & 0.0329 &  0.0500 & 0.5376 & 0.1346 &  0.0364 &  0.0585 & 0.5910$^{o}$\\
		\cline{2-10}
		& DAML & \textbf{0.1756} & 0.0596 &  0.086 & 0.5454 & 0.1789$^{o}$&  0.0618 &  0.0886$^{o}$ & 0.5876\\
		\cline{2-10}
		&	MetaCF & 0.1607 & 0.0746$^{o}$ & 0.1014$^{o}$ & 0.5383 & 0.1535 & 0.0807$^{o}$ & 0.0979 &  0.5024\\
		\cline{2-10}		
		&	 \textbf{MetaDPA } & 0.1702$^{o}$ & \textbf{0.0848} &  \textbf{0.1042} & 0.5803$^{o}$ & \textbf{0.2648} & \textbf{ 0.1371} & \textbf{ 0.1660}  & \textbf{0.6364 }	\\	
		\hline
		\hline
		
		\multirow{7}{*}{C-{I}}  & 
		NeuMF & 0.1248 & 0.0329 &  0.0536 & 0.5043 &  0.117 &  0.0416 &  0.0589 & 0.5054  \\
		\cline{2-10}
		&MeLU & 0.1846$^{o}$& 0.0610 &  0.0891$^{o}$ & 0.5977 & 0.1326 &  0.0404 &  0.0616 &  0.577 \\
		\cline{2-10}
		&CoNN & 0.1660 & 0.0438 &  0.0716 & 0.6014$^{o}$& 0.1871 &  0.0651 & 0.09319 & 0.6092 \\
		\cline{2-10}
		&	TDAR & 0.1275 & 0.0343 &  0.0556 & 0.5169 & 0.1423 &  0.0472 &  0.0688 & 0.5324\\
		\cline{2-10}
		&	CATN & 0.1280 & 0.0373 &  0.0581 & 0.5628 & 0.1195 &  0.0347 &  0.0539 & 0.5670\\
		\cline{2-10}
		&	DAML & 0.1766 & 0.0561 &  0.0839 & 0.5466 & 0.2008$^{o}$&  0.0738 &  0.1030 & 0.6143$^{o}$\\
		\cline{2-10}
		&	MetaCF & 0.1393 & 0.0631$^{o}$ & 0.0807 & 0.5298 & 0.1930 & 0.1041$^{o}$ & 0.1252$^{o}$ &  0.5108\\
		\cline{2-10}					
		&	\textbf{MetaDPA } & \textbf{0.2178} & \textbf{0.0775} & \textbf{ 0.1101 }& \textbf{0.6392 }& \textbf{0.3119} &  \textbf{0.1359 }&  \textbf{0.1762} & \textbf{0.6808}		\\
		\hline
		\hline
		
		\multirow{7}{*}{C-{UI}}  & 
		NeuMF & 0.0843 & 0.0277 &  0.0408 & 0.5387 & 0.1240 &  0.0371 &  0.0568 & 0.5144  \\
		\cline{2-10}
		&MeLU & 0.1742$^{o}$ & 0.0621$^{o}$ &  0.0872$^{o}$ & 0.5699$^{o}$ & 0.1240 &  0.0265 &  0.048  & 0.5733\\
		\cline{2-10}
		&CoNN & 0.1011 & 0.0223 &  0.0404 & 0.5493 & 0.1473 &  0.0570 &  0.0776 & 0.6045 \\
		\cline{2-10}
		&TDAR & 0.1180 & 0.0308 &  0.0504 & 0.5076 & 0.1705 &  0.0809 &  0.1016 & 0.5560\\
		\cline{2-10}
		&	CATN &  0.1209 & 0.0391 &  0.0579 & 0.5315 & 0.0916 &  0.0314 &  0.0451 & 0.5256\\
		\cline{2-10}
		&	DAML & 0.1292 & 0.0379 &  0.0590 & 0.5340 & 0.1550 &  0.0543 &  0.0774 & 0.6053$^{o}$\\
		\cline{2-10}
		&	MetaCF & 0.1236 & 0.0462 & 0.0644 & 0.5351 & 0.1860$^{o}$ & 0.0876$^{o}$ & 0.1104$^{o}$ &  0.5221 \\
		\cline{2-10}				
		&\textbf{	MetaDPA}  & \textbf{0.1910} & \textbf{0.0694 }&  \textbf{0.0971} & \textbf{0.5921 }&\textbf{ 0.2558} &  \textbf{0.1069} & \textbf{ 0.1407} & \textbf{0.6515}	\\
		\hline
		\hline
		
		\multirow{7}{*}{Warm-start}  
		& 	NeuMF & 0.1292 & 0.0415 &  0.0616 & 0.5107 & 0.1106 &  0.0336 &  0.0512 & 0.5014  \\
		\cline{2-10}
		&	MeLU & 0.1512 & \textbf{0.0929} &  0.1059 & 0.5562 & 0.1470 &  0.0465 &  0.0696 & 0.5826 \\
		\cline{2-10}
		&	CoNN & 0.1773 & 0.0528 &  0.0813 & 0.6023 & 0.1922 &  0.0645 &  0.0940 & 0.6029 \\ 
		\cline{2-10}
		&	TDAR & \textbf{0.2798} & 0.0749 & 0.1219$^{o}$ & \textbf{0.6331} & 0.1501 &  0.0489 &  0.0722 & 0.5843\\
		\cline{2-10}
		&	CATN &  0.1363 & 0.0386 & 0.0610  & 0.5539 & 0.1505 &  0.0440 &  0.0684 & 0.6183$^{o}$\\
		\cline{2-10}
		&	DAML &0.2011 & 0.0696 &  0.1000 & 0.5710 & 0.2007 & 0.0691 &  0.0995 & 0.6062 \\
		\cline{2-10}
		&	MetaCF & 0.1775 & 0.0857$^{o}$ & 0.1070 & 0.5127 & 0.2084$^{o}$ & 0.1511$^{o}$ & 0.1859$^{o}$ &  0.5472 \\
		\cline{2-10}				
		&	\textbf{MetaDPA } & 0.2099$^{o}$& 0.0829 &  \textbf{0.1396} & 0.6211$^{o}$ & \textbf{0.2866 }&  \textbf{0.1994} &  \textbf{0.2187} & \textbf{0.6481} \\
		\toprule
	\end{tabular}
\vspace{-14pt}
%	\label{tab:0}
\end{table*}	
			
	To investigate whether MetaDPA can improve the recommendation performance under cold-start settings or not, we compare it with several state-of-the-art baselines, including the competitive NeuMF, content-aware recommender systems (CoNN, DAML), meta-learning based methods (Melu, MetaCF), cross-domain recommendation frameworks (TDAR, CATN). 
	
	\begin{itemize}
		\item \textbf{NeuMF:} Neural collaborative filtering\cite{he2017neural2} is the most favorite technique in recommender systems. 
%		It is replacing the inner product of the latest features with a deep neural network structure. Benefiting from the strong representation of the deep learning technique, NCF shows satisfactory generalization ability when applied to learn any user-item interaction function.  
		\item \textbf{MeLU:} Meta-learned user preference estimator\cite{DBLP:conf/kdd/LeeIJCC19} demonstrates satisfactory performance when applied to a wide range of users for providing personalized recommendations due to the generalization ability.	
    	\item \textbf{MetaCF: } Fast adaptation for recommendations with meta-learning\cite{wei2020fast} is formulated with a dynamic subgraph sampling that accounts for the dynamic arrival of new users by dynamically generating representative adaptation tasks for existing users. 
%    	MetaCF optimizes the learning rates with a fine-grained way to learn a generalized adaption procedure that faces sparse ratings in cold-start settings.
		\item \textbf{CoNN:} Deep cooperative neural networks\cite{zheng2017joint} learns item properties and user behaviors from reviews consisting of two parallel neural networks with the shared last layer. 
%		One of the networks learns user behaviors, and the other one learns item properties. For a fair comparison, we feed the parallel neural networks with word semantic features extracted from text memory network (TMN) for users and items.
		\item \textbf{DAML:} Dual attention mutual learning\cite{liu2019daml} adapts local and mutual attention network to extract the rating and review features. With such features, neural factorization machine can effectively make predictions. 
%		For a fair comparison, the rating and review features are extracted from the semantic features prepossessed by the text memory network (TMN).
		\item \textbf{TDAR:} Text-enhanced domain adaptation recommendation (TDAR)\cite{DBLP:conf/kdd/YuLGOQ20} extracts the textual features in word semantic space for each user and item and feds them into a collaborative filtering model for predicting ratings. 
%		TDAR provides an effective textual feature extractor named Text Memory Network (TMN), which is used in our method. 
		\item \textbf{CATN:} Cross-domain recommendation framework via aspect transfer network\cite{zhao2020catn} learns cross-domain aspect-level preference matching by bridging multiple user’s inherent traits via reviews in different domains.
		\vspace{-10pt}
	\end{itemize}
	
	\begin{figure*}[!htbp]
	\centering
	\includegraphics[width=0.9\linewidth]{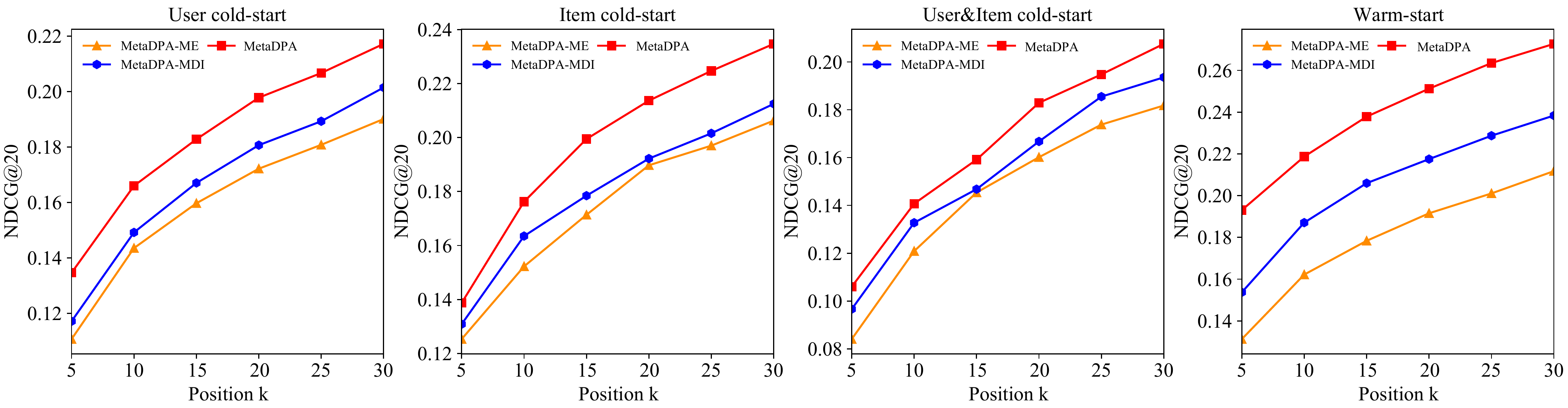}
			\vspace{-10pt}	
	\caption{Effectiveness of Mutually-Exclusive (ME) and Multi-domain InfoMax (MDI) constraints.}
	\vspace{-18pt}
	\label{fig_ablation}
\end{figure*}
%	\vspace{4pt}
	\subsubsection{Hyper-parameter Settings}
%	\vspace{4pt}
%	We have hyper-parameters $\beta_1$ for the MDI constraint imposed on the latent distributions for preserving domain-specific and domain-shared preference properties and $\beta_2$ for the ME constraint to force the generated ratings for the target domain close the source domain. 
	To obtain the optimal hyper-parameters $\beta_1$ and $\beta_2$ in Eq.\eqref{dual}, we apply the grid search in the range of $\{1e^{-2}, 1e^{-1},1, 1e^{1}, 1e^{2}\} \bigotimes \{1e^{-2}, 1e^{-1},1, 1e^{1}, 1e^{2}\}$ on Books and CDs, respectively. After searching, we set $\beta_1 = 0.1 $, $\beta_2 = 1$ on CDs, and $\beta_1 = 0.1 $, $\beta_2 = 1$ on Books, because these settings can achieve best performance.
	\vspace{-4pt}
	\subsection{Overall Performance Compared with Baselines (\textbf{RQ1 \& \textbf{RQ2}})}
%	\vspace{8pt}
	\label{sec_exp_over} 
	We compare our model MetaDPA with other competitive baselines to show the performance improvements under NDCG@k in Fig.~\ref{fig_books} and Fig.~\ref{fig_cds} on Books and CDs, respectively. The results under other metrics are shown in Table~\ref{performance}. We conclude that the performance of the proposed MetaDPA is significantly superior to all competing baselines. The overwhelming advantages over baselines attribute to the following three points: 
\vspace{-6pt}
	\begin{itemize}
		\item 	[(1)] Compared with cross-domain recommendations CATN and TDAR: By transferring preference properties from multiple source domains to the target domain via augmented diverse ratings, the proposed model can capture more preference properties, which is helpful to avoid overfitting and improve the recommendation accuracy, so it performs much better than cross-domain recommender systems; 
		\item	[(2)]Compared with content-aware recommendations CoNN and DAML: The diverse preference augmentation block aims to reduce the gap between content and preference that exists in content-aware recommender systems, so most of the time, it performs better than content-aware recommender systems; 
		\item [(3)] Compared with meta-learning based recommender systems Melu and MetaCF: With the diverse ratings fed into the preference meta-learning, MetaDPA avoids overfitting to the insufficient interactive training set, so it performs much better than meta-learning based recommender systems for solving recommendation problems.
	\end{itemize}

	\begin{figure}[!tbp]
	\centering
	\includegraphics[width=0.85\linewidth]{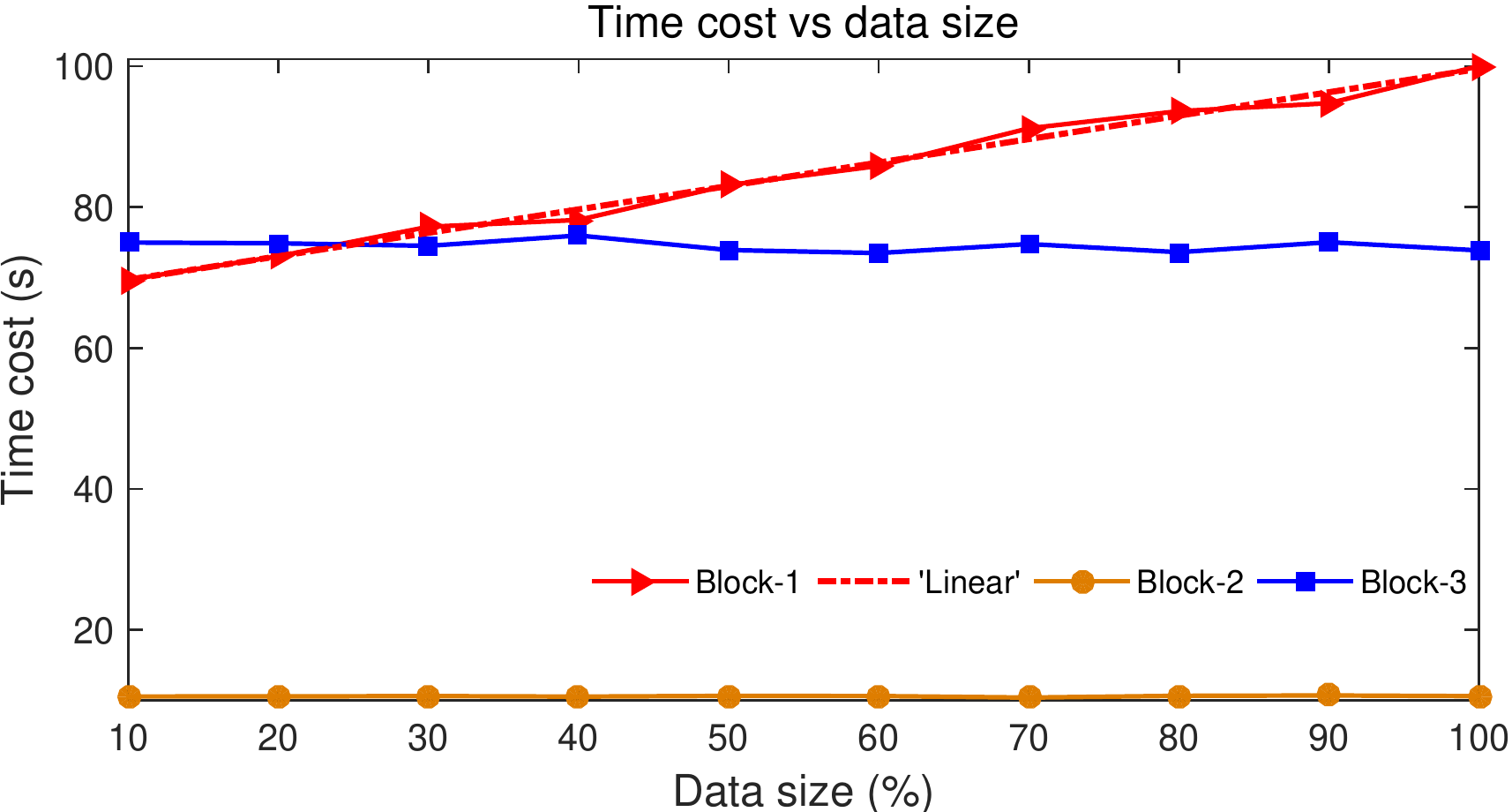}
	\vspace{-8pt}
	\caption{Training time cost varies with Data size.}
	\vspace{-20pt}
	\label{time}
\end{figure}

\begin{figure*}[!htbp]
	\centering
	\includegraphics[width=0.9\linewidth]{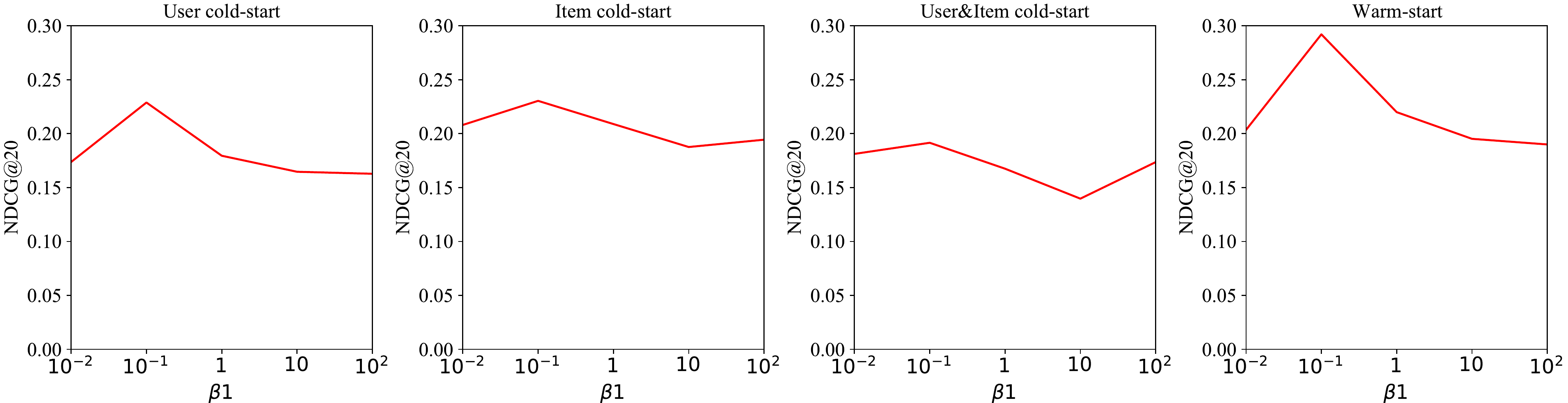}
	\vspace{-10pt}
	\caption{The impact of hyper-parameter $\beta_{1}$.}
		\vspace{-12pt}
	\label{fig_beta1}
\end{figure*}	
\begin{figure*}[!h]
	\centering
	\includegraphics[width=0.9\linewidth]{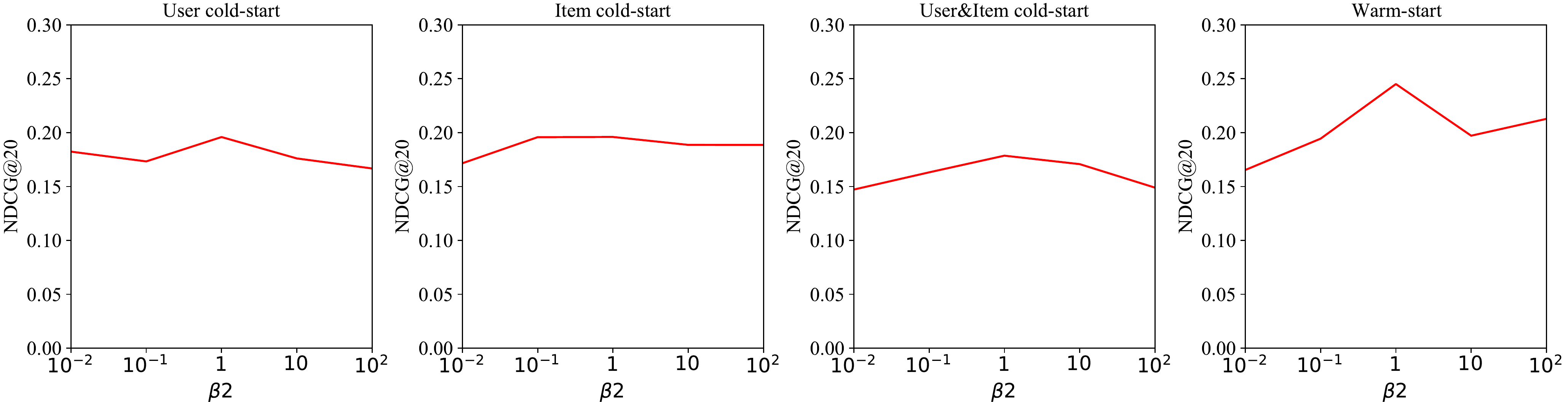}
	\vspace{-10pt}
	\caption{The impact of hyper-parameter $\beta_{2}$.}
		\vspace{-16pt}
	\label{fig_beta2}
\end{figure*}
%	Specifically, NCF is the deep version of the matrix factorization method by implementing with fully-connected layers to learn the latent factor of users and items. The competitive NCF is generic and widely used in various applications. However, in most scenarios of our experiments, it demonstrates the worst performance. The possible reason is that NCF is designed only with ratings, and it is common to suffer poor performance when the interactive data is sparse. 
	
	Specifically, both CoNN and DAML are content-aware recommendations with deep models. The difference is that CoNN uses a parallel neural network to learn user behaviors and item properties while DAML learns rating and review features by local and mutual attention networks. In our experiments, DAML shows a little better than CoNN. Both methods show middle performance in all scenarios. Actually, this indicates that content information can effectively deal with cold-start problems and work well in warm-start scenarios. 
	
%	Both TDAR and CATN are cross-domain content-aware recommendations by adopting different content feature extraction strategies. TDAR cares about semantic textual features for users and items. In contrast, CATN cares about aspect-level preference matching via reviews in different domains. 
  Besides, TDAR and CATN perform worse than content-based methods CoNN and DAML in most scenarios. However, the performance of TDAR is unstable. In some cases, it performs very well, e.g., the warm-start scenario on Books and CDs. The possible reason is that TDAR is designed for warm-start recommendations instead of cold-start settings, and the training datasets are very sparse for cold-start users/items, so it performs inferior to other baselines under cold-start settings and achieves better performance under warm-start settings. 
	
%	For fairness, we fuse the finetuning dataset(5 samples of each unseen user/item for finetuning the model introduced in Section V-A 2)) into the training dataset to train TDAR. As the training datasets are very sparse for cold-start users/items, so the performances of TDAR are inferior to other baselines under cold-start settings. However, for the warm-start, the result is superior to other baselines
	
%	It performs very bad and even obtains the worst performance on User cold-start and Item cold-start scenarios on Books.
		
	In addition, MeLU and MetaCF are effective meta-learning recommendations that are designed for solving cold-start issues with meta-learning frameworks. Both of them demonstrate powerful performances, which outperforms almost all other baselines in cold-start settings, and both perform well in the warm-start scenarios on Books as well. However, MeLU performs badly in all scenarios on CDs. A possible reason is that it easily sticks into the serious meta-overfitting on sparse user-item interactions. In contrast, MetaCF performs much better than other baselines on CDs, which indicates that incorporating potential interactions by neighborhood users / items can enrich users' preference information. 
		
	To explore the reasons why the performances of other baselines' on CDs are worse than the performances on Books, we calculate the proportion of users who have no less than 40 interactions, and we obtain 13\% for CDs and 16\% for Books, which may account for the different results of baselines. We are known that the recommendation performance is heavily dependent on the rating sparsity. When we learn models by a batch of users, the performances rely on the rating sparsity of a batch of users. If more users rated more items (such as more than 40 items), the recommendation problems would converges to a better solution, so the performances would be better.

%	We compare the proposed MetaDPA with the meta-learning based method MeLU under both cold-start and warm-start scenarios on CDs. MeLU meets a serious meta-overfitting problem. MetaDPA achieves significant performance improvement via the diverse preference augmentation for alleviating meta-overfitting. Especially for NDCG@5 in user\&items cold-start scenarios on CDs, MetaDPA outperforms MeLU 557.89\%. In addition, under both cold-start and warm-start settings on Books, MeLU does not meet a serious meta-overfitting problem. MetaDPA gains remarkable performance improvement. In the meta-learning stage, our method can be seen as MeLU with diverse tasks. This indicates the importance of diverse preference augmentation for meta-learning based recommendations for avoiding meta-overfitting. 

	Overall, under both cold-start and warm-start scenarios, MetaDPA outperforms all baselines significantly w.r.t NDCG@k metric. Regarding other metrics, MetaDPA outperforms almost all baselines except HR@10 and AUC for solving user cold-start problem (`C-U'), MRR@10 for solving `Warm-start' issue on Books. However, MetaDPA obtains the second-best performance on these problems. This indicates that meta-learning-based recommendations significantly improve cold-start recommendations and warm-start recommendations by addressing the meta-overfitting problem.
%	\vspace{-4pt}	
	
\subsection{Scalability (\textbf{RQ3})} 
   In this section, we run a series of experiments on the source domain Electronics and the target domain Books. To evaluate the scalability, we choose items in Books randomly with different percentages, 10\%, 20\%, ..., 100\% to create 10 group new datasets. To speed up the computation, we train our model on GPU platform (NVIDIA GeForce RTX 3090). In the experiments, we set batch size $B=32$ heuristically similar to the work\cite{DBLP:conf/www/YouWPERL19}, and we report the training time costs of 1 epoch in each block. As shown in Fig.~\ref{time}, the training procedure scales linearly with the data size for the first block (Block-1 in Fig.~\ref{time}), and the training time costs are constant with the data size for the second block (Block-2) and third block (Block-3). Which is consistent with the time analysis in Section \ref{sec-time}. So the proposed framework can be extended to larger datasets.

	\subsection{Significance Test}
   We use the one-sided test, Wilcoxon signed-rank test\cite{woolson2007wilcoxon}, to test the significance of our method surpassing the second-best methods, which has the null hypothesis that the median of the differences of two results $x_{i}-y_{i}$ under an evaluation metric ($x_{i}$ and $y_{i}$ denote the results of our method and the second-best method) is negative against the alternative that it is positive. For different metrics, we obtain different p-values. By randomly splitting training set and testing set 30 times independently, we obtain two sets of 30 results for our method and the second-best method, respectively.
	
By comparing with the second-best method MeLU, we test the significance on Books and get $p$-values are $5.96e^{-8}, 1.23e^{-5}, 1.78e^{-7}, 5.96e^{-8}$ for HR@10, MRR@10, NDCG@10, AUC under cold-start user (`C-U'). Similarly, for cold-start item (`C-I'), $p$-values are $5.96e^{-8}, 2.98e^{-7}, 5.96e^{-8}, 5.96e^{-8}$. For `C-UI' and `Warm-start', all $p$-values are $5.96e^{-8}$ except $p$-value is  $2.15e^{-4}$ under AUC for 'Warm-start' setting. Besides, we get all $p$-values are $1.19e^{-7}$ on CDs under all recommendation settings. Similarly, we obtain similar results when we compare the second-best methods MetaCF, DAML, TDAR, and CATN under specific settings. So we conclude that our method significantly outperforms other baselines with $p$-values<0.05 under all evaluation metrics.

%\begin{table}%[!t]
%	%\extrarowheight=3pt
%	
%	\caption{Statistics of datasets for target domains.}
%	% 		\small
%	\centering
%	%\begin{tabular*}{0.41\paperwidth}{@{\extracolsep{\fill}}ccccccc}
%	\begin{tabular}{c|c|c|c|c}
%		\toprule
%	\diagbox{Datasets}{$p$-value}	& \#users&\#items&\#ratings & sparsity \\
%		\hline
%		Books         & 603,374   &348,957  & 8,575,000  & 99.99\% \\
%		\hline
%		CDs         & 25,400  & 24,904 & 43,903    & 99.99\% \\
%		\toprule
%	\end{tabular}
%	\label{tab:target}
%\end{table}
%	\vspace{2pt}
	\subsection{Ablation Studies (\textbf{RQ4})}
%	\vspace{1pt}
	\label{sec_exp_abl}
	In this subsection, we discuss the effectiveness of MDI and ME constraints. We test two variants of MetaDPA to validate the effectiveness of the two constraints.
		\begin{itemize}
	 \item  \textbf{MetaDPA-ME}: MetaDPA only with the ME constraint. 
	 \item \textbf{MetaDPA-MDI}: MetaDPA only with the MDI constraint.
	\end{itemize}
%	\begin{itemize}
%	\item \textbf{MetaDPA-ME}: MetaDPA with only Mutually-exclusive (ME) constraint. 
%	\item \textbf{MetaDPA-MDI:} MetaDPA with only Multi-domain InfoMax (MDI) constraint.
%	\end{itemize}

	MetaDPA-ME only considers making the preference distribution of the target domain close to the source and ignores how to learn domain-specific and domain-shared preference properties for domain adaptation. MetaDPA-ME generates more diverse but less meaningful ratings for the target domain. In contrast, MetaDPA-MDI considers learning domain-specific and domain-shared properties as other cross-domain methods and ignores diverse requirements. The results on all four scenarios on CDs are shown in Fig.~\ref{fig_ablation}. Obviously, both MetaDPA-ME and MetaDPA-MDI suffer performance declines. MetaDPA-ME performs worst due to it generating meaningless ratings and introducing useless bias to the meta-training tasks. MetaDPA-MDI performs worse than MetaDPA because it generates ratings that are too close to the true ratings and lack diversity. However, both variants still perform better than MeLU and other baselines since the Dual-CVAE framework brings weak but good enough diversity and domain adaptation. 
	\vspace{-4pt}
	\subsection{The Impact of Hyper-parameters (\textbf{RQ5})} 
%	\vspace{2pt} 
	\label{sec_exp_rob} 
	In this subsection, we test the sensitivity of our framework w.r.t the hyper-parameters $\beta_1$ and $\beta_2$ on CDs by grid search introduced in Section \ref{sec_exp_set}. 
%	We apply a coarse liner search in the range of $\{1e^{-2}, 1e^{-1},1, 1e^{1}, 1e^{2}\}$ for both $\beta_1$ and $\beta_2$. 
The results for $\beta_1$ and $\beta_2$ are respectively shown in Fig.~\ref{fig_beta1} and Fig.~\ref{fig_beta2}. We conclude that $\beta_1$ is more sensitive than $\beta_2$. Actually, $\beta_1$ and $\beta_{2}$ weigh the importances of constraints MDI and ME, respectively. MDI affects both domain adaptation and diverse ratings generation; while ME only affects the latter, so the variation of $\beta_1$ affects the recommendation performance more seriously than $\beta_2$, i.e., $\beta_1$ is more sensitive than $\beta_2$. Besides, we find that warm-start scenarios are more sensitive than cold-start scenarios. As introduced in Section \ref{sec_exp_set}, we use a few samples to finetune the model (trained in the warm-start setting) for cold-start testings, so it has a better generalization ability than that for warm-start testing. That’s why warm-start scenarios are more sensitive than cold-start scenarios.

%	 However, even in the worst case, MetaDPA can still perform better than other baselines. This validates the robustness of MetaDPA to the hyper-parameters.
	
%	Besides, we use the grid search to choose the optimal β1 and β2 as introduced in Section V-A 4). Alternatively, we can also study the inherent function between the performances of the two constraints and then learn the optimal β1 and β2 automatically.
	
%	\vspace{-4pt}
	\section{Conclusion} \label{sec_con}
%	\vspace{-4pt}
	In this paper, we propose a Diverse Preference Augmentation based on meta-learning (MetaDPA) method by multiple source domains for solving cold-start issues in recommendation tasks. The proposed MetaDPA consists of three blocks. The first block is multi-source domain adaptation, formulated with multiple Dual-CVAEs. To preserve both domain-shared and domain-specific preference properties in the latent space, we add the MDI constraint with the principle of Info-Max, which is effective to maintain domain-shared properties without discarding domain-specific information. Besides, we impose the ME constraint on the outputs of decoders for generating diverse ratings from different source domains. The second block is diverse preference augmentation, which is realized by feeding the content data in the target domain into the content encoder $E_{t}^{x}$ and the decoder $D_{t}$. The last block, preference meta-learning, optimizes the preference model which is formulated with a multi-layer neural network based on the model-agnostic meta-learning scheme. Experimental results clearly show that MetaDPA significantly outperforms state-of-the-art baselines including content-aware, cross-domain, and meta-learning methods on public datasets. Besides, we conduct ablation studies to evaluate the effectiveness of two constraints MDI and ME. Finally, we also evaluate the impacts of hyper-parameters $\beta_{1}$ and $\beta_{2}$ on the recommendation performance.
\section*{Acknowledgments}
	\vspace{0pt}
This work is supported by the Major Project for New Generation of AI under Grant No. 2018AAA0100400, the National Natural Science Foundation of China (Grant No. 62002052, 61801060, 62176047, 62172075, 61832001), the Sichuan Science and Technology Program (Grant No. 2022YFG0189, 2021JDRC0079). Besides, this work is supported in part by Australian Research Council (Grant No. FT210100624, DP180100106, DP200101328, DP190101985), the Beijing Natural Science Foundation (Grant No. Z190023), the China Postdoctoral Science Foundation (Grant No. 2019TQ0051), the Open Fund of Intelligent Terminal Key Laboratory of Sichuan Province (Grant No. SCITLAB-1017).

	\balance
	\bibliographystyle{IEEEtran}
	\bibliography{reference}

\end{document}